\documentclass[fleqn,usenatbib]{mnras}

\usepackage[T1]{fontenc}
\usepackage{ae,aecompl}

\usepackage{times} % for details on the use of the package, please
\usepackage{tikz}
\usetikzlibrary{positioning}
\usepackage{xspace,ulem}

\usetikzlibrary{backgrounds}
\usetikzlibrary{arrows}
\usetikzlibrary{calc}
\usepackage{booktabs}
\usepackage{graphicx}	% Including figure files
\usepackage{amsmath}	% Advanced maths commands
\usepackage{amssymb}	% Extra maths symbols
\usepackage{times}
\usepackage{newtxtext}
\usepackage[varvw]{newtxmath}

\title[Realistic cluster simulations in modified gravity]{A general framework to test gravity using galaxy clusters VI: \\ Realistic galaxy formation simulations to study clusters in modified gravity}

\author[M.~A.~Mitchell et al.]{
Myles A.~Mitchell,\thanks{E-mail: m.a.mitchell@durham.ac.uk}
Christian Arnold
and Baojiu Li
\\
Institute for Computational Cosmology, Department of Physics, Durham University, South Road, Durham DH1 3LE, UK
}

\date{Accepted XXX. Received YYY; in original form ZZZ}

\pubyear{2022}

\begin{document}
\label{firstpage}
\pagerange{\pageref{firstpage}--\pageref{lastpage}}
\maketitle

\begin{abstract}
We present a retuning of the IllustrisTNG baryonic physics model which can be used to run large-box realistic cosmological simulations with a lower resolution. This new model employs a lowered gas density threshold for star formation and reduced energy releases by stellar and black hole feedback. These changes ensure that our simulations can produce sufficient star formation to closely match the observed stellar and gas properties of galaxies and galaxy clusters, despite having $\sim160$ times lower mass resolution than the simulations used to tune the fiducial IllustrisTNG model. Using the retuned model, we have simulated Hu-Sawicki $f(R)$ gravity within a $301.75h^{-1}{\rm Mpc}$ box. This is, to date, the largest simulation that incorporates both screened modified gravity and full baryonic physics, offering a large sample ($\sim$500) of galaxy clusters and $\sim$8000 galaxy groups. We have reanalysed the effects of the $f(R)$ fifth force on the scaling relations between the cluster mass and four observable proxies: the mass-weighted gas temperature, the Compton $Y$-parameter of the thermal Sunyaev-Zel'dovich effect, the X-ray analogue of the $Y$-parameter, and the X-ray luminosity. We show that a set of mappings between the $f(R)$ scaling relations and their $\Lambda$CDM counterpart, which have been tested in a previous work using a much smaller cosmological volume, are accurate to within a few percent for the $Y$-parameters and $\lesssim7\%$ for the gas temperature for cluster-sized haloes ($10^{14}M_{\odot}\lesssim M_{500}\lesssim10^{15}M_{\odot}$). These mappings will be important for unbiased constraints of gravity using the data from ongoing and upcoming cluster surveys.
\end{abstract}


%%%%%%%%%%%%%%%%%%%%%%%%%%%%%%%%%%%%%%%%%%%%%%%%%%

%%%%%%%%%%%%%%%%% BODY OF PAPER %%%%%%%%%%%%%%%%%%

\section{Introduction}
\label{sec:introduction}

Galaxy clusters are the largest gravitationally-bound structures that have been observed in the Universe. They are believed to trace the highest peaks of the primordial density field, and their properties are highly sensitive to cosmological parameters which control the formation and evolution of large-scale cosmological structure. For example, the cluster abundance, which is quantified using cluster number counts, has been used to constrain the matter density parameter $\Omega_{\rm M}$ and the linear fluctuation of the density field $\sigma_8$ \citep[e.g.,][]{Planck_SZ_cluster}. Clusters can also be used to probe the strength of gravity on the largest scales: for example, a strengthened gravity would result in the formation of a greater number of clusters by the present-day and alter properties such as the temperature of the intra-cluster gas. Therefore, cluster observations can be used to search for departures from General Relativity (GR).

Ongoing and upcoming astronomical surveys are generating vast catalogues using all available means of detection: clustering of galaxies in galaxy surveys \citep[e.g.,][]{ukidss,lsst,euclid,desi}; X-ray emission produced by the hot intra-cluster gas \citep[][]{chandra,xmm-newton,erosita}; and the Sunyaev-Zel'dovich (SZ) effect \citep[e.g.,][]{1972CoASP...4..173S}, which is caused by interactions between cosmic microwave background photons and electrons in the intra-cluster gas \citep[e.g.,][]{act,Planck_SZ_cluster,Abazajian:2016yjj,Ade:2018sbj}. Some of these catalogues will be many times larger than previous data sets, and will offer a means to significantly enhance our understanding of the mechanisms driving the large-scale evolution of the Universe, including the late-time accelerated cosmic expansion.

Before we can use this data to test cosmological models, we must first consider any potential sources of bias that can arise from an incomplete modelling of the cluster properties. For example, many modified gravity (MG) models \citep[see, e.g.,][]{Clifton:2011jh,Joyce:2015PhR...568....1J,Koyama:2020zce,Ferreira:2019xrr}, which can be used to explain the accelerated cosmic expansion, feature a strengthened gravitational force. This can alter properties such as the intra-cluster gas temperature and the density profile. If these effects are not properly accounted for in tests of these models using the abundance and other properties of galaxy clusters, the inferred constraints may be subject to bias. For example, the cluster mass is often determined using scaling relations that relate the mass $M$ with some observable flux $Y_{\rm obs}$. In GR, these are often modelled as power laws, but this may be inaccurate for MG. If these effects are not accounted for, then constraints inferred by comparing the observed cluster abundance, ${\rm d}n/{\rm d}Y_{\rm obs}$, with the model-dependent prediction of the halo mass function (HMF), ${\rm d}n/{\rm d}M$, can be unreliable, since the inferred cluster mass is not correct.

The above effects can be studied and modelled using numerical cosmological simulations which incorporate sub-resolution models for baryonic processes such as star formation, gas cooling, and stellar and black hole feedback \citep[e.g.,][]{Schaye:2014tpa,2017MNRAS.465.3291W,Pillepich:2017jle}. For the first time, these `full-physics' models are being incorporated in MG simulations to study the combined effects of the extra gravitational forces and baryonic processes on the properties of galaxies. For example, the \textsc{shybone} simulations \citep[see][]{Arnold:2019vpg,Hernandez-Aguayo:2020kgq}, which incorporate the IllustrisTNG baryonic physics model \citep{2017MNRAS.465.3291W,Pillepich:2017jle}, were run for two popular MG models which feature a strengthened gravity {and some screening mechanism to recover GR in high-density regions}. These yielded useful insights into, for example, the abundance of disk-shaped galaxies, the power spectra of different matter components, and the stellar and gas properties of galaxies in these models. {As an example, it was found that, for MG models with weak or moderate deviations from GR, the impacts of baryons and MG can be modelled separately and added up, which means that full baryonic simulations in MG are not necessary for predicting the matter power spectrum on small scales. Some interesting observations were also made about the number of disk galaxies produced by different models \citep[][]{Arnold:2019vpg}, with a stronger gravity leading to more frequent halo mergers and therefore fewer surviving disk galaxies (though this observation requires further verification using larger boxes). The $f(R)$ simulation data also show that different gravity models could produce very different numbers of small haloes (those less massive than $10^{10}M_\odot$), and hence lead to different amounts and clustering of 21cm-emitting neutral hydrogen \citep[][]{Leo:2019ada}. But those simulations are generally aimed at studies of galaxy properties, and so they have very high resolution and small box sizes which are not suitable for a cosmological study.} Extending this approach from the galactic regime to the cluster regime is now an important step as we prepare for future cluster tests of gravity.

The $f(R)$ gravity model \citep[e.g.,][]{Sotiriou:2008rp,DeFelice:2010aj} is a popular MG model that can give rise to the accelerated expansion without violating local Solar System tests. The model is a modification of GR, which introduces an extra scalar field that couples to matter, giving rise to a `fifth force' which enhances the total strength of gravity. Past works have used various cluster properties to probe $f(R)$ gravity, including: number counts \citep[e.g.,][]{PhysRevD.92.044009,Liu:2016xes,Peirone:2016wca}; the clustering of clusters \citep{Arnalte-Mur:2016alq}; redshift-space distortions \citep[e.g.,][]{Bose:2017dtl,2018NatAs...2..967H,Hernandez-Aguayo:2018oxg,Garcia-Farieta:2021hda}; the gas mass fraction \citep[e.g.,][]{Li:2015rva}; the temperature-mass relation \citep[see, e.g.,][]{Hammami:2016npf,DelPopolo:2019oxn}; the SZ profile \citep{deMartino:2016xso}; comparisons of weak lensing data with thermal observables \citep[e.g.,][]{Terukina:2013eqa,Wilcox:2015kna}; and the angular power spectra of the thermal and kinetic SZ effects \citep[e.g.,][]{Ma:2013taq,Bianchini:2015iaa,Roncarelli:2018kud,Mitchell:2020fnj}. 

This paper is part of a series of works which aims to develop a general framework for robust and unbiased tests of gravity using galaxy clusters. An important component of this framework is a mapping %between
of observable-mass scaling relations %in
from GR %and
to $f(R)$ gravity, to accurately predict the latter from the (better-known knowledge of the) former. As mentioned in the above, this is required in order to avoid biased mass estimates. Most interestingly, this can be %evaluated
done using a simple analytical model for the $f(R)$ enhancement of the dynamical mass --- in \citet{Mitchell:2020aep}, we tested this mapping using the \textsc{shybone} simulations, and demonstrated that it performs very well for galaxy groups. However, the \textsc{shybone} simulations have a box size of just $62h^{-1}{\rm Mpc}$ (`L62') and a high mass resolution, %. This 
which as mentioned above is more suited for studying galaxies than galaxy clusters; indeed, these simulations contain only a few cluster-sized objects ($M_{500}\gtrsim10^{14}M_{\odot}$)\footnote{In this work, we define the halo mass $M_{\Delta}$ as the total mass within a sphere of radius $R_{\Delta}$ which is centred on the potential minimum of the halo and encloses an average density of $\Delta$ times the critical density at the halo redshift for a flat universe.}. The L62 predictions of the cluster scaling relations may therefore suffer from poor statistics and be potentially subjected to a significant influence by sample variance. On the other hand, it has been shown that a sufficiently high mass resolution is required for the IllustrisTNG model to generate sufficient levels of star formation to closely match galaxy observations \citep[e.g.,][]{Pillepich:2017jle}. Unfortunately, high-resolution simulations which incorporate both screened modified gravity and full baryonic physics are very expensive to run for larger cosmological volumes, which has made it difficult to study the interplay between baryonic physics and the fifth force at higher masses. 

In this work, we will present a retuning of the IllustrisTNG model which can be used to run full-physics simulations at lower resolutions without losing the good agreement with observational data. This retuning was a significant undertaking which involved running over 200 simulations
with a reduced box size, and our new model can be used to run low-resolution, large-box simulations for both standard gravity and MG scenarios. We have used this model to run simulations for GR and $f(R)$ gravity with a significantly increased box size of $301.75h^{-1}{\rm Mpc}$, and in this work we will present the predictions for the observable-mass scaling relations over an extended mass range $10^{13}M_{\odot}\leq M_{500}\lesssim10^{15}M_{\odot}$.

This paper is structured as follows. In Sec.~\ref{sec:background}, we provide some background on the $f(R)$ gravity model and our general framework. Then, in Sec.~\ref{sec:methods}, we provide a detailed description of the baryonic physics retuning and the large-box simulations, including the agreement with galaxy observations. We present our results for the observable-mass scaling relations in Sec.~\ref{sec:results}. Finally, we provide a summary of this paper in Sec.~\ref{sec:conclusions}.

\section{Background}
\label{sec:background}

In Sec.~\ref{sec:background:theory}, we will introduce the $f(R)$ gravity model studied in this work. Then, in Sec.~\ref{sec:background:clusters}, we will describe the effects of the $f(R)$ fifth force on the properties of galaxy clusters and outline our general framework. Throughout this section, we will adopt the unit convention $c=1$ for the speed of light and Greek indices can take values 0, 1, 2 and 3. Unless stated otherwise, overbars (e.g., $\bar{x}$) will be used to denote the background value of a quantity, while a subscript $_0$ will denote the present-day value of a quantity.

\subsection{Theory}
\label{sec:background:theory}

The $f(R)$ gravity model is constructed by adding a nonlinear function, $f(R)$, of the Ricci scalar curvature, $R$, to the integrand of the Einstein-Hilbert action of GR:
\begin{equation}
    S=\int {\rm d}^4x\sqrt{-g}\left[\frac{R+f(R)}{16\pi G}+\mathcal{L}_{\rm M}\right],
\label{eq:action}
\end{equation}
where $g$ is the determinant of the metric tensor $g_{\alpha\beta}$, $G$ is Newton's constant and $\mathcal{L}_{\rm M}$ is the Lagrangian density for matter fields (we will focus on late-time cosmology and therefore on non-relativistic matter). By setting the variation of the action to zero, the modified Einstein field equations can be derived:
\begin{equation}
    G_{\alpha \beta} + X_{\alpha \beta} = 8\pi GT_{\alpha \beta},
\label{eq:modified_field_equations}
\end{equation}
where $G_{\alpha\beta}$ is the Einstein tensor, $T_{\alpha\beta}$ is the stress-energy tensor and $X_{\alpha\beta}$ is a new tensor which is given by:
\begin{equation}
    X_{\alpha \beta} = f_RR_{\alpha \beta} - \left(\frac{f}{2}-\Box f_R\right)g_{\alpha \beta} - \nabla_{\alpha}\nabla_{\beta}f_R,
\label{eq:GR_modification}
\end{equation}
where $R_{\alpha\beta}$ is the Riemann tensor, $\Box\equiv\nabla_{\alpha}\nabla^{\alpha}$ is the d'Alembert operator and $\nabla_{\alpha}$ is the covariant derivative compatible with the metric $g_{\alpha\beta}$. 

The quantity $f_R\equiv{\rm d}f/{\rm d}R$ represents the extra scalar degree of freedom which we refer to as the `scalar field'. This mediates a fifth force which can act on scales smaller than the Compton wavelength of the scalar field:
\begin{equation}
    \lambda_{\rm C} = a^{-1}\left(3\frac{{\rm d}f_R}{{\rm d}R}\right)^{\frac{1}{2}},
\label{eq:compton_wavelength}
\end{equation}
where $a$ is the cosmological scale factor. The chameleon screening mechanism \citep[e.g.,][]{Khoury:2003aq,Khoury:2003rn,Mota:2006fz} is featured by the model in order to suppress the fifth force in high-density regions, ensuring consistency with Solar System tests of gravity \citep[e.g.,][]{Will:2014kxa}. This is brought about by an effective mass of the scalar field, $m_{\rm eff}=\lambda^{-1}_{\rm C}$, which becomes very large in dense regions, significantly reducing the range of the fifth force. The fifth force will therefore only act in regions with shallow gravitational potential, which can include cosmic voids, low-mass haloes and the outer regions of galaxy clusters. In these regions, the total strength of gravity is enhanced by up to a factor of $4/3$.

In this work, we will focus on the popular Hu-Sawicki (HS) model of $f(R)$ gravity \citep{Hu:2007nk}, which uses the following prescription for the function $f(R)$:
\begin{equation}
    f(R) = -m^2\frac{c_1\left(-R/m^2\right)^n}{c_2\left(-R/m^2\right)^n+1},
\label{eq:hu_sawicki}
\end{equation}
where $m^2\equiv8\pi G\bar{\rho}_{\rm M,0}/3=H_0^2\Omega_{\rm M}$, with $\bar{\rho}_{\rm M,0}$ the present-day background matter density and $H_0$ the Hubble constant. The model has three free parameters: $n$, $c_1$ and $c_2$. However, it can be simplified by assuming $-\bar{R}\gg m^2$ for the background curvature, in which case the background scalar field is given by:  
\begin{equation}
\bar{f_R} \approx -n\frac{c_1}{c_2^2}\left(\frac{m^2}{-\bar{R}}\right)^{n+1}.
\label{eq:scalar_field}
\end{equation}
Assuming that the background expansion history is practicably indistinguishable from $\Lambda$CDM, the background curvature is given by:
\begin{equation}
-\bar{R} = 3m^2\left(a^{-3}+4\frac{\Omega_\Lambda}{\Omega_{\rm M}}\right),
\label{eq:R}
\end{equation}
where $\Omega_{\Lambda}=1-\Omega_{\rm M}$. Therefore, the background scalar field at a given redshift $z$ can be re-expressed as:
\begin{equation}
    \bar{f}_R(z) = f_{R0}\left(\frac{\bar{R}_0}{\bar{R}(z)}\right)^{n+1},
\end{equation}
where $f_{R0}$ is the present-day value of the background scalar field (we will omit the overbar from this quantity throughout this work). We note that the inequality $-\bar{R}\gg m^2$ holds for a realistic choice of cosmological parameter values. Therefore, to a reasonable accuracy we are able to work with just two free parameters: $n$ and $f_{R0}$. We will choose $n=1$ throughout this work, which is also a common choice in literature, and we will be working with a value $|f_{R0}|=10^{-5}$ (the `F5' model). The amplitude of the background scalar field is greater at later times, which means that a halo with a given mass is more likely to be unscreened (i.e., it can feel the fifth force) at late times.

\subsection{\boldmath Galaxy clusters in \texorpdfstring{$f(R)$}{f(R)} gravity}
\label{sec:background:clusters}

%%%%%%%%%%%%%%%%%%%%%%%%%%%%%%%%%%%%%%%%%%%%%%%%%%
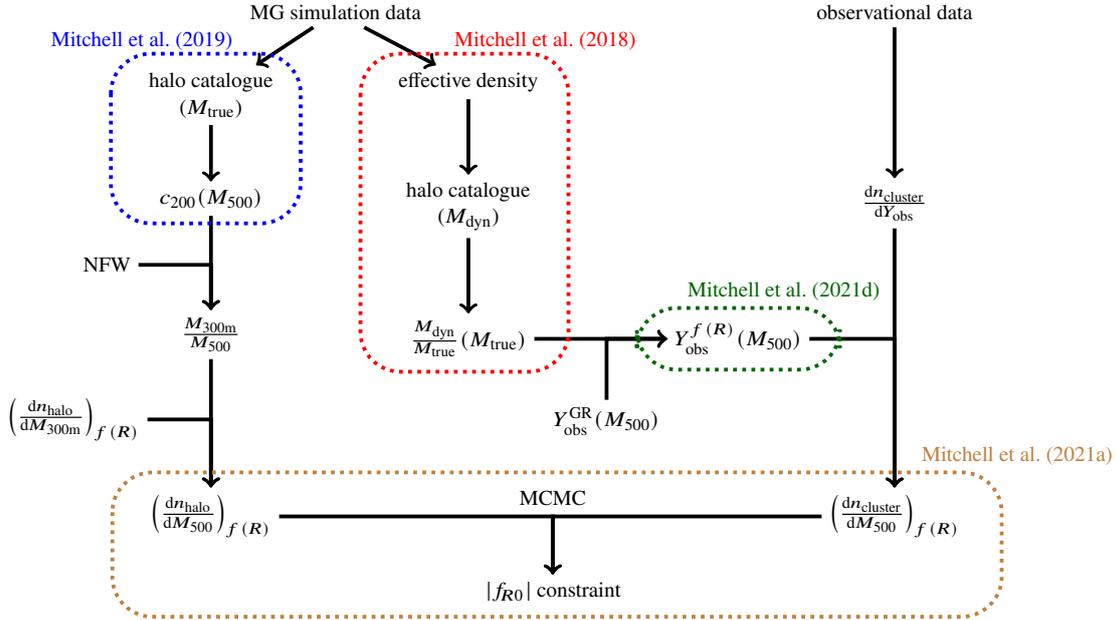
\begin{figure*}
\centering
\begin{tikzpicture}
\tikzstyle{myarrow}=[line width=0.5mm,draw=black,-triangle 45,postaction={draw, line width=0.5mm, shorten >=4mm, -}]

%Nodes
%Simulation Side
\node    (simulations)    {MG simulation data};
%left branch
\node    (cat_true)    [below left = 0.5cm and -0.5cm of simulations]   [align=center]{halo catalogue \\ ($M_{\rm true}$)};
\node    (c_m)    [below = 0.75cm of cat_true] [align=center]   {$c_{\rm 200}(M_{\rm 500})$};
\node   (m300_m500)    [below = 1.25cm of c_m] [align=center]    {$\frac{M_{\rm 300m}}{M_{500}}$};
\node    (nfw)    [below left = 0.65cm and 1.0cm of c_m]  [anchor=west]   {NFW};
\node    (hmf_m300)    [below left = 2.7cm and 2.0cm of c_m]  [anchor=west]   {$\left( \frac{{\rm d} n_{\rm halo}}{{\rm d}M_{\rm 300m}} \right)_{f(R)}$};
\node    (hmf_th)    [below = 1.7cm of m300_m500]  [align=center]   {$\left( \frac{{\rm d} n_{\rm halo}}{{\rm d} M_{500}} \right)_{f(R)}$};
%right branch
\node    (rho_eff)    [below right = 0.5cm and -0.5cm of simulations]   [align=center]  {effective density};
\node    (cat_mdyn)    [below = 1.0cm of rho_eff]    [align=center] [align=center]{halo catalogue \\ ($M_{\rm dyn}$)};
\node    (mdyn_mtrue)    [below = 1.0cm of cat_mdyn]  [align=center]   {$\frac{M_{\rm dyn}}{M_{\rm true}}(M_{\rm true})$};
%\node    (hmf_mdyn_th)    [below = 1.75cm of mdyn_mtrue]  [align=center]   {$\left( \frac{{\rm d} n_{\rm halo}}{{\rm d} M_{\rm 500,dyn}} \right)_{f(R)}$};
%Observation Side
\node    (observations)    [right = 5cm of simulations]    [align=center] {observational data};
\node    (n_Y)    [below = 2.0cm of observations]   [align=center]  {$\frac{{\rm d} n_{\rm cluster}}{{\rm d}Y_{\rm{obs}}}$};
\node    (hmf_obs)  at (hmf_th -| n_Y)  [ align=center]   {$\left(\frac{{\rm d} n_{\rm cluster}}{{\rm d} M_{\rm 500}}\right)_{f(R)}$};
\node    (scaling_relation)    [right = 1.75cm of mdyn_mtrue]    [align=center]{$Y_{\rm{obs}}^{f(R)}(M_{500})$};
\node    (scaling_relation_lcdm)    [below left= 0.5cm and -0cm of scaling_relation]    [align=center]{$Y_{\rm{obs}}^{\rm GR}(M_{500})$};

%Combination and Results
\node    (mcmc)    at ($(hmf_th)!0.5!(hmf_obs)+(0.0,0.25)$)   [align=center]  {MCMC};
\node    (constraint)  at ($(hmf_th)!0.5!(hmf_obs)+(0.0,-1.0)$) [align=center]  {$|f_{R0}|$ constraint};

%Lines
%left
\draw[->, line width=0.5mm] (simulations) -- (cat_true);
\draw[->, line width=0.5mm] (cat_true) -- (c_m);
\draw[->, line width=0.5mm] (c_m) -- (m300_m500);
\draw[->, line width=0.5mm, to path={-| (\tikztotarget)}] (nfw) edge (m300_m500);
\draw[->, line width=0.5mm] (m300_m500) -- (hmf_th);
\draw[->, line width=0.5mm, to path={-| (\tikztotarget)}] (hmf_m300) edge (hmf_th);
%right
\draw[->, line width=0.5mm] (simulations) -- (rho_eff);
\draw[->, line width=0.5mm] (rho_eff) -- (cat_mdyn);
\draw[->, line width=0.5mm] (cat_mdyn) -- (mdyn_mtrue);

%obs
\draw[->, line width=0.5mm] (observations) -- (n_Y);
\draw[->, line width=0.5mm] (n_Y) -- (hmf_obs);
\draw[->, line width=0.5mm, to path={-| (\tikztotarget)}] (scaling_relation) edge (hmf_obs);
\draw[->, line width=0.5mm] (mdyn_mtrue) -- (scaling_relation);
\draw[->, line width=0.5mm, to path={|- (\tikztotarget)}] (scaling_relation_lcdm) edge (scaling_relation);

%join
\draw[->, line width=0.5mm, to path={-| (\tikztotarget)}] (hmf_th) edge (constraint);
\draw[->, line width=0.5mm, to path={-| (\tikztotarget)}] (hmf_obs) edge (constraint);

%\draw[->, line width=0.5mm] (hmf_th) -- (mcmc);
%\draw[->, line width=0.5mm] (hmf_mdyn_obs) -- (mcmc);
%\draw[->, line width=0.5mm] (mcmc) -- (constraint);

%Projects
\draw [line width=0.5mm,dotted, red, rounded corners=15pt]     ($(rho_eff.north west)+(-0.4,0.15)$) rectangle ($(mdyn_mtrue.south east)+(0.45,-0.1)$);
\node [above right = 0.10cm and -1.3cm of rho_eff] {\small{\color{red}\hypersetup{citecolor=red}\cite{Mitchell:2018qrg}}}; 
\draw[line width=0.5mm,dotted, blue, rounded corners=15pt]   ($(cat_true.north west)+(-0.4,0.15)$) rectangle ($(c_m.south east)+(0.4,-0.1)$);
\node [above left = 0.10cm and -1.3cm of cat_true] {\small{\color{blue}\hypersetup{citecolor=blue}\cite{Mitchell:2019qke}}}; 
\draw[line width=0.5mm,dotted, brown, rounded corners=15pt]     ($(hmf_th.north west)+(-0.4,0.2)$) rectangle ($(constraint.south east -| hmf_obs.south east) +(0.45,-0.1)$);
\node [above right = 0.20cm and -0.7cm of hmf_obs] {\small{\color{brown}\hypersetup{citecolor=brown}\cite{Mitchell:2021uzh}}}; 
\draw[line width=0.5mm,dotted, black!60!green, rounded corners=15pt]     ($(scaling_relation.north west)+(-0.4,0.1)$) rectangle ($(scaling_relation.south east) +(0.4,-0.1)$);
\node [above right = 0.10cm and -1.7cm of scaling_relation] {\small{\color{black!60!green}\hypersetup{citecolor=black!60!green}\cite{Mitchell:2020aep}}}; 
\end{tikzpicture}
\caption{[{\it Colour Online}] Flowchart illustrating our framework to constrain $f(R)$ gravity using galaxy cluster number counts. In \citet{Mitchell:2019qke} and \citet{Mitchell:2018qrg}, we used dark-matter-only simulations to calibrate models for the $f(R)$ enhancements of the halo concentration (\textit{blue dotted box}) and the dynamical mass (\textit{red dotted box}). Our model for the concentration is required for conversions between halo mass definitions. In \citet{Mitchell:2020aep}, we showed that our model for the dynamical mass enhancement can be used to convert observable-mass scaling relations from GR to $f(R)$ gravity (\textit{green dotted box}). These can be used to link the observed cluster abundance to the theoretical halo mass function. In \citet{Mitchell:2021uzh}, we used mock cluster catalogues to validate our MCMC pipeline for constraining the amplitude of the present-day background scalar field (\textit{brown dotted box}). This pipeline will be employed in future works to test gravity.}
\label{fig:flow_chart}
\end{figure*}
%%%%%%%%%%%%%%%%%%%%%%%%%%%%%%%%%%%%%%%%%%%%%%%%%%

In this subsection, we describe the main components of our framework to test gravity using galaxy clusters, as illustrated in Fig.~\ref{fig:flow_chart}, including our previous work on cluster scaling relations in $f(R)$ gravity.

\subsubsection{Dynamical mass enhancement}

In this work, we will refer to two definitions of the cluster mass (or the mass of any massive body whose gravity is of interest): the `true' mass is the intrinsic mass, and can be inferred using weak lensing; the `dynamical' mass is the mass that is felt by a nearby massive test particle, and can be measured using properties related to the total gravitational potential of the halo, including the velocity dispersion and thermal properties including the X-ray temperature. In GR, the two masses are expected to be equal: $M_{\rm true}^{\rm GR}=M_{\rm dyn}^{\rm GR}$ (we will therefore neglect the subscript from the GR mass, $M^{\rm GR}$, in what follows); while in $f(R)$ gravity the fifth force will enhance the dynamical mass relative to the true mass: $M_{\rm true}^{f(R)}\leq M_{\rm dyn}^{f(R)}\leq (4/3)M_{\rm true}^{f(R)}$. In \citet{Mitchell:2018qrg}, we used dark-matter-only (DMO) simulations to calibrate a general model for the ratio of the dynamical mass to the true mass in HS $f(R)$ gravity. This is accurately described by a tanh fitting formula, with the dynamical mass of low-mass (unscreened) haloes enhanced by a factor of 4/3 and with no enhancement for high-mass haloes which are efficiently screened:
\begin{equation}
\frac{M^{f(R)}_{\rm dyn}}{M^{f(R)}_{\rm true}} = \frac{7}{6}-\frac{1}{6}\tanh\left(p_1\left[\log_{10}\left(M^{f(R)}_{\rm true}M_{\odot}^{-1}h\right)-p_2\right]\right).
\label{eq:mdyn_enhancement}
\end{equation}
We found that the parameter $p_1$ is approximately constant, with a best-fit value of $p_1=2.21\pm0.01$. For the parameter $p_2$, we obtained the following best-fit relation:
\begin{equation}
    p_2=(1.503\pm0.006)\log_{10}\left(\frac{|\bar{f}_R(z)|}{1+z}\right)+(21.64\pm0.03).
\label{eq:p2}
\end{equation}
This parameter represents the logarithmic halo mass above (below) which haloes are expected to be mainly screened (unscreened). Our model, which is an important component of the framework (cf.~the red dotted box in Fig.~\ref{fig:flow_chart}), attains a very high accuracy for a wide range of halo masses ($10^{11}h^{-1}M_{\odot}\lesssim M_{500}\lesssim10^{15.5}h^{-1}M_{\odot}$) and present-day strengths of the scalar field ($10^{-6.5}\leq|f_{R0}|\leq10^{-4}$).

\subsubsection{Halo concentration}

In \citet{Mitchell:2019qke}, we used an extended suite of DMO simulations to calibrate a general model for the enhancement of the halo concentration in $f(R)$ gravity (blue dotted box in Fig.~\ref{fig:flow_chart}). The concentration is an important parameter of the Navarro-Frenk-White density profile of dark matter haloes \citep{NFW}. Our model can therefore be used to model the effect of the fifth force on the halo density profile, allowing for conversions between halo masses defined with respect to different spherical overdensities. We will not provide the model here, since it is not required for the present study.

\subsubsection{Observable-mass scaling relations}
\label{sec:background:clusters:scaling_relations}

It is often difficult and (observationally) expensive to directly measure the dynamical mass of clusters. This can require long exposure times and high-quality spectra and X-ray data. Instead, the cluster mass is often inferred using its one-to-one relationship with the thermal properties of the intra-cluster gas. During cluster formation, the initial gravitational potential energy of the gas is converted into thermal energy through shock heating as it is accreted by the dark matter halo. In GR, this leads to an approximate power-law mapping between the cluster mass and various thermal properties (`mass proxies') such as the gas temperature $T_{\rm gas}$, the Compton $Y$-parameter of the SZ effect $Y_{\rm SZ}$, the X-ray analogue of the $Y$-parameter $Y_{\rm X}$ and the X-ray luminosity $L_{\rm X}$. These observable-mass scaling relations have been widely studied in the literature using both hydrodynamical simulations \citep[e.g.,][]{Brun:2016jtk,Truong:2016egq} and observations \citep[e.g.,][]{Ade:2013lmv}. In $f(R)$ gravity, the gravitational potential of a halo can be enhanced by up to $1/3$ by the fifth force, and consequently the gas temperature may be enhanced by a similar factor, leading to departures from a power-law \citep[e.g.,][]{He:2015mva,Hammami:2016npf} because the departure will depend on the mass of the halo (stronger for low-mass haloes).

In \citet{Mitchell:2020aep}, we tested two methods for mapping between the observable-mass scaling relations in GR and $f(R)$ gravity using the full-physics \textsc{shybone} simulations (the green dotted box in Fig.~\ref{fig:flow_chart}). One of these methods (the `effective density' approach) had already been proposed and studied using non-radiative simulations by \citet{He:2015mva}. We also proposed and tested an alternative set of mappings (the `true density' approach). We found that both methods work well for galaxy groups and low-mass clusters for the $Y$-parameters and the gas temperature, even in the presence of the extra baryonic processes found in the full-physics simulations, including star formation, cooling and feedback. However, the \textsc{shybone} simulations only contain haloes with mass $M_{500}\lesssim10^{14.5}M_{\odot}$, including only a few cluster-sized objects ($M_{500}\gtrsim10^{14}M_{\odot}$). In this work, we will be using larger simulations to verify the previous results with a much larger halo sample and an extended mass range $10^{13}M_{\odot}\leq M_{500}\lesssim10^{15}M_{\odot}$. We will be focusing on the `true density' mappings, which are summarised below.

Consider a halo in GR which has mass $M^{\rm GR}$ and gas temperature $T_{\rm gas}^{\rm GR}$, and a halo in $f(R)$ gravity which has true mass $M_{\rm true}^{f(R)}$, dynamical mass $M_{\rm dyn}^{f(R)}$ and gas temperature $T_{\rm gas}^{f(R)}$. If the true masses of these haloes are equal (i.e., if $M^{\rm GR}=M_{\rm true}^{f(R)}$), then the following relation between their gas temperatures is expected:
\begin{equation}
    T^{f(R)}_{\rm gas}\left(M^{f(R)}_{\rm true}\right) = \frac{M_{\rm dyn}^{f(R)}}{M_{\rm true}^{f(R)}}T^{\rm GR}_{\rm gas}\left(M^{\rm GR}=M^{f(R)}_{\rm true}\right).
    \label{eq:temp_equiv_true}
\end{equation}
This relation can be understood as follows: as discussed above, the gas temperature is closely related to the total gravitational potential, which is approximately $GM_{\rm dyn}/R_{500}$. Because the dynamical mass in $f(R)$ gravity can be enhanced by up to 1/3 by the fifth force, the gas temperature of the $f(R)$ halo is expected to be enhanced relative to the temperature of the GR halo by the same factor\footnote{Note that for haloes in GR and $f(R)$ gravity of equal true mass $M_{\rm true}\equiv M_{500}$, the mass definition guarantees that their radius $R_{500}$ is the same too.}, giving rise to the $M_{\rm dyn}^{f(R)}/M_{\rm true}^{f(R)}$ factor in Eq.~(\ref{eq:temp_equiv_true}). This factor also arises in the relations between the $f(R)$ and GR $Y$-parameters, since these are linearly related to the gas temperature:
\begin{equation}
    Y_{\rm SZ}^{f(R)}\left(M_{\rm true}^{f(R)}\right) \approx \frac{M_{\rm dyn}^{f(R)}}{M_{\rm true}^{f(R)}}Y_{\rm SZ}^{\rm GR}\left(M^{\rm GR}=M_{\rm true}^{f(R)}\right),
    \label{eq:ysz_mapping_true}
\end{equation}
\begin{equation}
    Y_{\rm X}^{f(R)}\left(M_{\rm true}^{f(R)}\right) \approx \frac{M_{\rm dyn}^{f(R)}}{M_{\rm true}^{f(R)}}Y_{\rm X}^{\rm GR}\left(M^{\rm GR}=M_{\rm true}^{f(R)}\right).
    \label{eq:yx_mapping_true}
\end{equation}
On the other hand, the X-ray luminosity %is
varies as $T_{\rm gas}^{1/2}$, therefore the $f(R)$ and GR values are related by a factor of $\left(M_{\rm dyn}^{f(R)}/M_{\rm true}^{f(R)}\right)^{1/2}$:
\begin{equation}
    L_{\rm X}^{f(R)}\left(M_{\rm true}^{f(R)}\right) \approx \left(\frac{M_{\rm dyn}^{f(R)}}{M_{\rm true}^{f(R)}}\right)^{1/2}L_{\rm X}^{\rm GR}\left(M^{\rm GR}=M_{\rm true}^{f(R)}\right).
    \label{eq:lx_mapping_true}
\end{equation}
The $Y$-parameters and the X-ray luminosity also depend on the gas density. During the formation of a cluster, matter is drawn from an initially large region, and so the ratio between the gas mass and total mass is expected to be equal to the cosmic baryonic fraction $\Omega_{\rm b}/\Omega_{\rm M}$ to very good approximation \citep[e.g.,][]{1993Natur.366..429W}. Therefore, two haloes with the same true mass are expected to have a similar gas content regardless of whether a fifth force is acting or not, and so the gas density dependence of the above observables does not contribute substantial additional factors in Eqs.~(\ref{eq:ysz_mapping_true})-(\ref{eq:lx_mapping_true}). In \citet{Mitchell:2020aep}, we validated this assumption for group-sized haloes using the \textsc{shybone} simulations (see Fig.~2 in that work), and showed that the mappings given by Eqs.~(\ref{eq:temp_equiv_true})-(\ref{eq:lx_mapping_true}) can be computed analytically using Eq.~(\ref{eq:mdyn_enhancement}) for the mass ratio.

\subsubsection{Further works}

In a recent work \citep{Mitchell:2021uzh}, we tested our general framework by using Markov chain Monte Carlo (MCMC) sampling to constrain the present-day amplitude of the background scalar field, $|f_{R0}|$, using mock cluster data (brown dotted box in Fig.~\ref{fig:flow_chart}). In future works, we plan to use this sampling pipeline to constrain $|f_{R0}|$ using real cluster data from ongoing and upcoming surveys. Our pipeline is also designed to be easily extendable to other MG models and cluster observables. For example, we recently \citep{Mitchell:2021aex} carried out a similar modelling of the cluster properties in the normal-branch Dvali-Gabadadze-Porrati (nDGP) model of gravity \citep{DVALI2000208}, and we have also demonstrated the potential in using the thermal and kinetic SZ angular power spectra to probe $f(R)$ gravity and nDGP \citep{Mitchell:2020fnj}. Both of these works were carried out using the L62 \textsc{shybone} simulations, therefore it will be important to validate these results using simulations with much larger box sizes before we use our pipeline to test gravity using real data. This will be another use for the new baryonic model (see Sec.~\ref{sec:methods:fine_tuning}), which can be used to run large-box simulations for a range of MG scenarios (in addition to GR).

\section{Simulations and methods}
\label{sec:methods}

In Secs.~\ref{sec:methods:illustristng} and \ref{sec:methods:fine_tuning}, we will provide a brief summary of the IllustrisTNG model and discuss our retuning of the model parameters for lower-resolution simulations. Our large-box simulations, which are used for the main results of this paper, are presented in Sec.~\ref{sec:methods:L302_simulations}.

\subsection{The IllustrisTNG model}
\label{sec:methods:illustristng}

In this section, we will briefly summarise the main features of the IllustrisTNG subgrid model \citep{2017MNRAS.465.3291W,Pillepich:2017jle} and the $N$-body and hydrodynamical simulation code \textsc{arepo} \citep{2010MNRAS.401..791S}, where this model is implemented. The description will be kept concise, with further relevant details given in the next subsection.

In \textsc{arepo}, dark matter and gas are respectively sampled as particles and cells. The gas cells in \textsc{arepo} make up an unstructured, moving Voronoi mesh. The cells are adaptive in the way that they refine (split) and derefine (merge), such that the mass of any cell does not differ by more than a factor of two from the mean. The code uses a tree-particle-mesh algorithm to solve the Poisson equation and a second order finite-volume Godunov scheme on the Voronoi mesh to solve the ideal magneto-hydrodynamics equations, with Powell cleaning used to maintain the divergence constraint of the magnetic field \citep[see][]{Pakmor2011, Pakmor2013}. The magnetic field, which dynamically couples to gas through magnetic pressure, is initially seeded at $z=127$ with a uniform strength of $1.6\times10^{-10}~{\rm Gauss}$.  

The TNG model employs a subgrid scheme for star formation which is based on the \citet{Springel:2002uv} model: at each simulation timestep, for gas cells which exceed a particular threshold density, a fraction of the gas mass is converted to mass in star particles according to the Kennicutt-Schmidt law \citep{Pillepich:2017jle}. A star particle represents a population of stars with an initial mass function given by \citet{Chabrier:2003ki}. The evolution of these stars and the subsequent chemical enrichment of the surrounding gas is tracked. A portion of the gas mass in star-forming gas cells is also converted into wind particles which are launched in random directions \citep{Pillepich:2017jle}; these represent galactic winds driven by supernovae. These wind particles will eventually couple to gas cells outside their local dense interstellar medium, resulting in the heating and metal enrichment of the gas. Gas cells also undergo radiative cooling which is modulated by a time-dependent ultraviolet background radiation.

Supermassive black holes are seeded at the centre of friends-of-friends (FOF) groups which exceed a particular mass threshold. These can then grow through a combination of Eddington-limited Bondi gas accretion and black hole mergers. The TNG model employs two types of black hole feedback, depending on the accretion state of the black hole \citep{2017MNRAS.465.3291W, Vogelsberger2013}: in the low accretion state, a kinetic feedback model is employed which produces black hole-driven winds; while in the high accretion state, a thermal feedback model is employed which heats up the surrounding gas. 

It has been shown that a sufficiently high mass resolution is required in order to use the IllustrisTNG model to accurately reproduce the observed properties of galaxy populations \citep[e.g.,][]{Pillepich:2017jle}. For example, a lowered mass resolution means that the gas cells will have larger volumes, resulting in a smoothed out density field which can miss out the high-density peaks where star formation would be highest. The L62 \textsc{shybone} simulations, with $512^3$ gas cells, have $\sim$15 times lower mass resolution than that used to calibrate the TNG model ($25h^{-1}{\rm Mpc}$ box with the same number of gas cells). The gas and stellar properties of haloes from the L62 simulations still give a reasonable agreement with observational data, however the lowered resolution means that there is less star formation, resulting in the amplitudes of the stellar mass fraction, the stellar mass function and the star formation rate density (SFRD) being reduced compared to the fiducial TNG results \citep[for example, see Fig.~4 in][]{Arnold:2019vpg}.

\subsection{Baryonic physics fine-tuning}
\label{sec:methods:fine_tuning}

For the present work, running $f(R)$ gravity simulations with a substantial number of galaxy clusters (with masses $10^{14}M_{\odot}\lesssim M_{500}\lesssim10^{15.5}M_{\odot}$) requires a box size of at least $\sim300h^{-1}{\rm Mpc}$, which necessitates going to even lower resolutions than the L62 simulations to remain computationally feasible. In order to make this possible without losing the good agreement with observational data, it is necessary to retune the parameters of the IllustrisTNG model, including parameters which control the density threshold for star formation and the energy released by the stellar and black hole feedback mechanisms.

To recalibrate the baryonic physics at our desired mass resolution, we have run a large number of realisations with a small box size of $68h^{-1}{\rm Mpc}$ (`L68-N256'). These runs have $256^3$ dark matter particles with mass $1.35\times10^9h^{-1}M_{\odot}$ and, initially, the same number of gas cells with mass $\sim2.6\times10^{8}h^{-1}M_{\odot}$ on average. The calibration was carried out using GR (although, as we will show, the retuned model works equally well for F5) with the same cosmological parameter values as the \textsc{shybone} simulations: ($h$, $\Omega_{\rm M}$, $\Omega_{\rm b}$, $n_{\rm s}$, $\sigma_8$) $=$ ($0.6774$, $0.3089$, $0.0486$, $0.9667$, $0.8159$), where $h=H_0/(100{\rm kms^{-1}Mpc^{-1}})$ and $n_{\rm s}$ is the power-law index of the primordial power spectrum. The runs were all started from the same set of initial conditions at redshift $z=127$. These have been generated using the \textsc{N-GenIC} code \citep[e.g.,][]{Springel:2005nw}, which uses the Zel’dovich approximation to displace an initially homogeneous particle distribution and obtain an initial density field with a prescribed linear power spectrum. Each of the input particles is then split into a dark matter particle and a gas cell, with the ratio of masses set by the values of the cosmic density parameters $\Omega_{\rm M}$ and $\Omega_{\rm b}$.

%%%%%%%%%%%%%%%%%%%%%%%%%%%%%%%%%%%%%%%%%%%%%%%%%
\begin{figure*}
\centering
\includegraphics[width=0.87\textwidth]{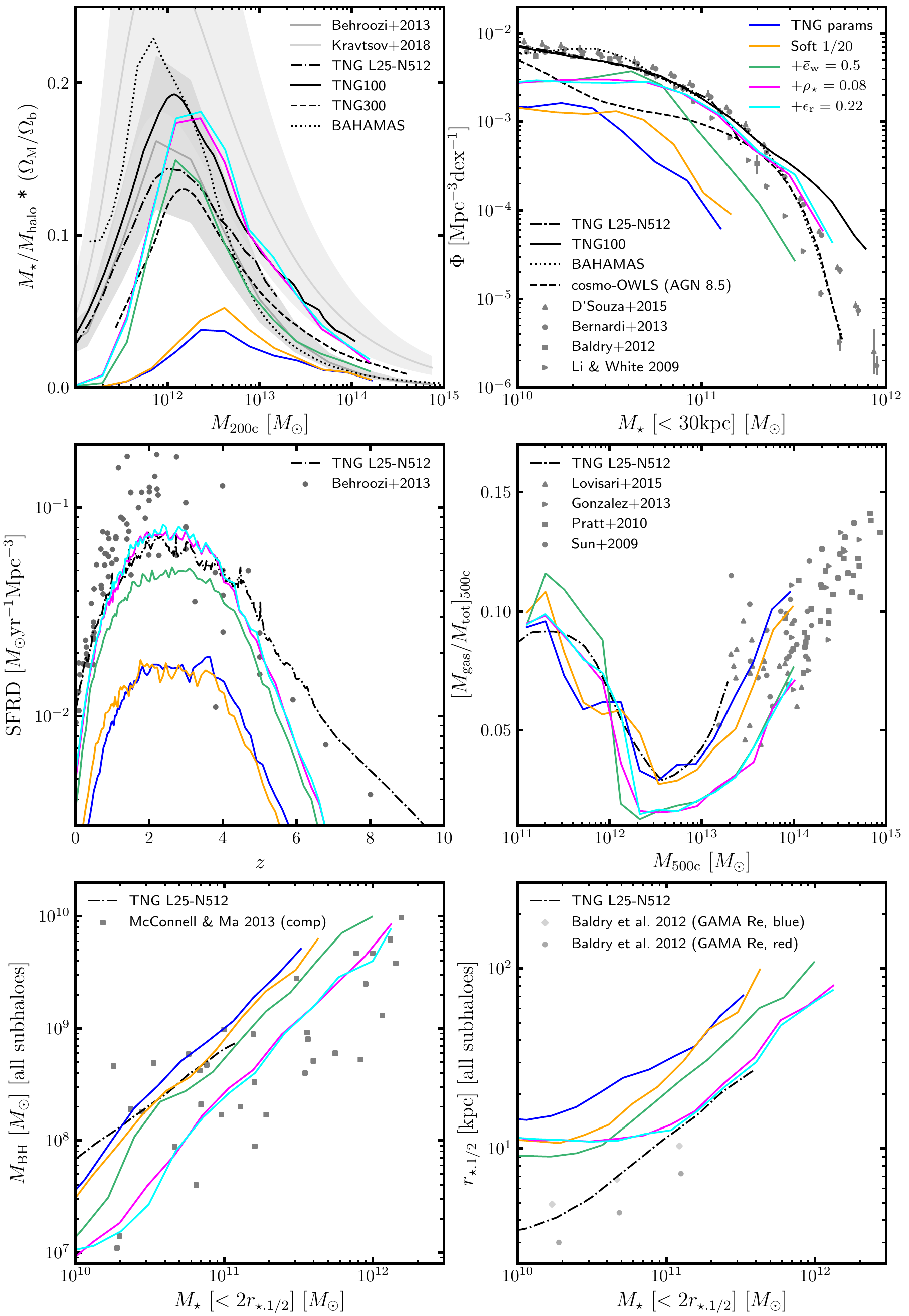}
\caption{[{\it Colour Online}] Stellar, gas and black hole properties in a sample of the L68-N256 calibration runs (\textit{coloured solid lines}). The properties are (\textit{clockwise from top-left}): stellar mass fraction; stellar mass function; gas mass fraction; stellar half-mass radius; black hole mass; star formation rate density. A selection of simulation (black lines with different styles) and observational (symbols or shaded regions) results from the literature are shown as a comparison. See the legends and Sec.~\ref{sec:methods:fine_tuning} for further details.}
\label{fig:L68_calibration}
\end{figure*}
%%%%%%%%%%%%%%%%%%%%%%%%%%%%%%%%%%%%%%%%%%%%%%%%%%

Halo catalogues are constructed using the \textsc{subfind} code \citep{springel2001} which is implemented in \textsc{arepo}. This uses the FOF algorithm to identify FOF groups (haloes) and a gravitational un-binding method to locate the bound substructures (subhaloes) of each group. By adjusting the baryonic physics parameters of our calibration runs, we have aimed for reasonable agreement with observational data and empirical constraints for the six galaxy properties shown in Fig.~\ref{fig:L68_calibration}, which were also used to calibrate the IllustrisTNG model \citep{Pillepich:2017jle}. These are: the stellar mass fraction (within halo radius $R_{\rm 200c}$), with empirical constraints from \citet{2013ApJ...770...57B} and \citet{Kravtsov:2014sra}; the stellar mass function (subhaloes), with observations from \citet{2015MNRAS.454.4027D}, \citet{Bernardi:2013mqa}, \citet{2012MNRAS.421..621B} and \citet{2009MNRAS.398.2177L}; the SFRD as a function of redshift, with observations from \citet{2013ApJ...770...57B}; the gas mass fraction (within halo radius $R_{\rm 500c}$), with observations from \citet{Lovisari:2014pka}, \citet{Gonzalez:2013awy}, \citet{2010A&A...511A..85P} and \citet{Sun:2008eh}; the black hole mass versus the stellar mass (subhaloes), with the compilation of observations from \citet{2013ApJ...764..184M}; and the galaxy size versus the stellar mass (subhaloes), with observational data from \citet{2012MNRAS.421..621B}. The results for a selection of our calibration runs are represented by the coloured solid lines. Apart from the SFRD, which is a direct output of the simulations, these lines are generated using mass-binning of either haloes or subhaloes (see the parentheses above). The black lines show predictions from the TNG simulations \citep[e.g.,][]{Nelson:2017cxy,Springel:2017tpz,Marinacci:2017wew,Pillepich:2017fcc,2018MNRAS.477.1206N} as well as the BAHAMAS and cosmo-OWLS simulations \citep{McCarthy:2016mry,Brun:2013yva}.

The dark blue line in Fig.~\ref{fig:L68_calibration} shows the predictions using the fiducial TNG model at our lowered resolution. Star formation is significantly reduced at this resolution compared to the fiducial TNG resolution, which is used by the `TNG L25-N512' simulation (`TNG100' has a similar resolution, while `TNG300' has $\sim5$ times lower resolution). Consequently, the stellar mass fraction, the stellar mass function and the SFRD are significantly lower. The primary objective of our retuning is therefore to achieve a greater amount of star formation in order to obtain a closer match with the observational data. Our changes are described in the sections below, and the effects of these changes are shown in Fig.~\ref{fig:L68_calibration}. We note that the calibration runs discussed in this section are only a very small subset of the $\sim$200 simulations which were run for this calibration study: we provide further details of these simulations and the calibration procedure in Appendix \ref{appendix:baryonic_fine_tuning}.

\subsubsection{Gravitational softening}

In low-resolution simulations, where the gas cells have higher masses, there is a higher risk of two-body heating: this occurs when two particles come close together and incur a significant gravitational boost to their velocities, which, if happening frequently, can raise the internal energy and subsequently the temperature of the gas. We have therefore increased the gravitational softening length to $1/20$ times the mean inter-particle separation, which is about twice the length used for the \textsc{shybone} simulations. The gravitational force is dampened when gas cells come within this distance, preventing extreme interactions. This change alone causes an overall reduction of the gas temperature in our simulations, which results in more cool gas that is capable of forming stars: see the orange lines in Fig.~\ref{fig:L68_calibration}, which have a greater amplitude than the dark blue lines for the stellar mass fraction and stellar mass function. 

We also considered fractions of $1/30$ and $1/10$ for the softening length. For higher-mass haloes ($M_{200}\gtrsim10^{12}M_{\odot}$), we observed that using a larger softening length results in greater star formation (for the reasons discussed above). However, we were unable to significantly boost star formation at lower masses; in fact, we observed that a large softening length (for example, a fraction 1/10 of the mean inter-particle separation) can even lead to less star formation for low-mass haloes. A potential effect of using a larger softening is that the gravitational potential well of haloes effectively becomes shallower. For low-mass haloes, where the gravitational potential well is already shallower than for high-mass haloes, this could potentially lead to a lower density of cold gas (e.g., the gas is now less gravitationally bound) which in turn could reduce star formation. This is a motivation for using the fraction 1/20, for which we never observed the above effect, rather than using larger fractions.

It is evident from Fig.~\ref{fig:L68_calibration} that, while it can increase star formation, changing the gravitational softening length alone is not enough to produce stellar contents that match observational data.

\subsubsection{Stellar feedback}
\label{sec:fine_tuning:stellar_feedback}

For a star-forming gas cell with metallicity $Z$, the available wind energy is \citep{Pillepich:2017jle}:
\begin{equation}
\begin{split}
    e_{\rm w} = &~\bar{e}_{\rm w}\left[f_{{\rm w},Z}+\frac{1 - f_{{\rm w},Z}}{1 + (Z/Z_{\rm w,ref})^{\gamma_{{\rm w},Z}}}\right]\\
    &\times N_{\rm SNII}E_{\rm SN11,51}10^{51}{\rm erg}M_{\odot}^{-1},
\end{split}
\label{eq:wind_energy}
\end{equation}
where $\bar{e}_{\rm w}$ is a dimensionless free parameter, $E_{\rm SNII,51}$ is the available energy from core-collapse supernovae in units of $10^{51}{\rm erg}$, $N_{\rm SNII}$ is the number of supernovae per stellar mass that is formed, and $f_{{\rm w},Z}$, $Z_{\rm w,ref}$ and $\gamma_{{\rm w},Z}$ are additional parameters of the model. A wind particle will eventually donate its thermal energy (along with its mass, momentum, and metal content) to a gas cell that is outside its local dense inter-stellar medium. This heats the gas and subsequently reduces the efficiency of star formation (gas must be sufficiently cool in order to form stars). 

Star formation efficiency is reduced by our lowered gas cell resolution, therefore reducing the thermal heating of the gas by wind feedback can help to rectify this. We have achieved this in our retuning of the model by reducing the value of $\bar{e}_{\rm w}$ from the fiducial TNG value 3.6 to 0.5, which lowers the energy of the winds. As can be seen from the green lines in Fig.~\ref{fig:L68_calibration}, this change significantly boosts star formation over a wide range of halo masses. The stellar mass fraction now has a reasonable amplitude for $M_{200}\gtrsim10^{12}M_{\odot}$, while the amplitudes of the SFRD and stellar mass function are much closer to the observational data.

We have also tried varying the wind speed $v_{\rm w}$, which in the IllustrisTNG model is given by \citep{Pillepich:2017jle}:
\begin{equation}
    v_{\rm w} = {\rm max}\left[\kappa_{\rm w}\sigma_{\rm DM}\left(\frac{H_0}{H(z)}\right)^{1/3},v_{\rm w,min}\right],
\label{eq:wind_speed}
\end{equation}
where $\kappa_{\rm w}$ is a dimensionless factor, $\sigma_{\rm DM}$ is the local one-dimensional velocity dispersion of the dark matter particles and $v_{\rm w,min}$ is the minimum wind velocity allowed in the model. For our calibration runs, we tried reducing the $\kappa_{\rm w}$ and $v_{\rm w,min}$ parameters. This reduces the speed of the wind particles, which now take longer to transfer the thermal energy to the surrounding gas. Gas is therefore heated up at a slower rate, resulting in an increased amount of star formation. We found that reducing these parameters has a similar effect to reducing the $\bar{e}_{\rm w}$ parameter, with star formation boosted over a wide halo mass range. 

However, we could find no clear advantage in varying the wind speed parameters instead of $\bar{e}_{\rm w}$, or in varying all three parameters in combination. Moreover, it has been impossible to boost star formation sufficiently to get the stellar mass function and fraction matching observational data by varying these parameters. For simplicity, we therefore decided to adjust the stellar feedback using only the $\bar{e}_{\rm w}$ parameter.

\subsubsection{Star formation model}
\label{sec:methods:fine_tuning:star_formation_model}

As discussed in Sec.~\ref{sec:methods:illustristng}, the star formation rate in IllustrisTNG is computed for gas cells using the \citet{Springel:2002uv} model. Stars can only be formed by gas cells which exceed a particular density threshold, which is approximately $n_{\rm H}\approx0.1{\rm cm}^{-3}$. We will refer to the value of the threshold gas density as $\rho_{\star}$ in this work. At our reduced resolution, gas cells have a larger volume and therefore a smoothed density which can miss out high-density peaks of cold gas in galaxies. In order to account for this, we have reduced $\rho_{\star}$ from $\approx0.1$ to a fixed value of 0.08, allowing gas cells with lower density to form stars. 

The effect of making this change, in addition to the changes listed above, is shown by the magenta lines in Fig.~\ref{fig:L68_calibration}. This further boosts the stellar mass fraction and SFRD, which are now both in good agreement with the TNG100 results for $M_{200}\gtrsim10^{12}M_{\odot}$ and the TNG L25-N512 results for $z\lesssim5$, respectively, and there is also now a good agreement with the \citet{2015MNRAS.454.4027D}, \citet{Bernardi:2013mqa} and \citet{2012MNRAS.421..621B} observations of the stellar mass function for $M_{\star}\gtrsim10^{11}M_{\odot}$.

\subsubsection{Black hole feedback}
\label{sec:methods:fine_tuning:bh_feedback}

In IllustrisTNG, the rate of gas accretion, $\dot{M}$, by the central supermassive black holes is set by the Eddington-limited Bondi accretion rate \citep{2017MNRAS.465.3291W}:
\begin{equation}
\begin{split}
    &\dot{M}_{\rm Bondi}=\frac{4\pi G^2M_{\rm BH}^2\rho}{c_{\rm s}^3},\\
    &\dot{M}_{\rm Edd}=\frac{4\pi GM_{\rm BH}m_{\rm p}}{\epsilon_{\rm r}\sigma_{\rm T}c},\\
    &\dot{M}={\rm min}\left(\dot{M}_{\rm Bondi},\dot{M}_{\rm Edd}\right),
\end{split}
\label{eq:accretion_rate}
\end{equation}
where $M_{\rm BH}$ is the black hole mass, $\rho$ represents the ambient density around the black hole, $c_{\rm s}$ is the ambient sound speed and $\epsilon_{\rm r}$ is the black hole radiative efficiency. The mode of feedback depends on whether or not the ratio $\dot{M}/\dot{M}_{\rm Edd}$ exceeds the following threshold:
\begin{equation}
    \chi = {\rm min}\left[\chi_0\left(\frac{M_{\rm BH}}{10^8M_{\odot}}\right)^{\beta},0.1\right],
\end{equation}
where $\chi_0$ and $\beta$ are parameters. If $\dot{M}/\dot{M}_{\rm Edd}>\chi$, the resulting thermal feedback will inject thermal energy into the surrounding gas at a rate $\dot{E}_{\rm therm}=\epsilon_{\rm f,high}\epsilon_{\rm r}\dot{M}c^2$, where $\epsilon_{\rm f,high}$ is another parameter; and if $\dot{M}/\dot{M}_{\rm Edd}<\chi$, the resulting kinetic feedback will inject energy into the surroundings at a rate $\dot{E}_{\rm kin}=\epsilon_{\rm f,kin}\dot{M}c^2$, where the factor $\epsilon_{\rm f,kin}$ depends on the ambient density $\rho$. Both of these feedback modes will reduce the efficiency of star formation in the surrounding gas, either by blowing gas out, so that less gas will exceed the density threshold for star formation, or by heating up gas which, as for stellar feedback, reduces the amount of cool gas capable of forming stars. As discussed above, the star formation efficiency is already reduced by our lowered gas resolution; reducing the overall effect of black hole feedback on star formation therefore provides another means of rectifying this.

For our retuning of the black hole feedback, we have increased $\epsilon_{\rm r}$ from the fiducial TNG value $0.2$ to $0.22$. The effect of this change on the overall energy release is quite complex: the energy injected by thermal feedback will be boosted, unless $\dot{M}=\dot{M}_{\rm Edd}$ (i.e., $\dot{M}_{\rm Bondi}>\dot{M}_{\rm Edd}$) in which case the $\epsilon_{\rm r}$ factors will cancel and there will be no effect; on the other hand, from Eq.~(\ref{eq:accretion_rate}) we see that $\dot{M}_{\rm Edd}$ is lowered if $\epsilon_{\rm r}$ is increased, and subsequently the ratio $\dot{M}/\dot{M}_{\rm Edd}$ will be greater and there will then be less kinetic feedback. From this discussion, increasing $\epsilon_{\rm r}$ is therefore expected to increase the heating of the gas by thermal feedback and reduce the blowing out of gas by kinetic feedback: two effects which would have competing impacts on the star formation efficiency. For our calibration runs, we have observed that increasing $\epsilon_{\rm r}$ from $0.18$ to $0.22$ slightly boosts the amount of star formation. Therefore, it seems that the reduced blowing out of gas by kinetic feedback has the dominant effect here.

The result of making this final adjustment to the baryonic physics model is shown by the cyan lines in Fig.~\ref{fig:L68_calibration}. The stellar mass fraction and stellar mass function are both slightly boosted for high-mass haloes. From the upper-right panel of Fig.~\ref{fig:L68_calibration}, our model now appears to slightly overshoot the observed stellar mass function at higher masses; this is actually a consequence of sample variance which results from using a small box-size. As we will show in Sec.~\ref{sec:methods:L302_simulations}, the agreement is very good for the %much 
larger $301.75h^{-1}{\rm Mpc}$ box size. The change to $\epsilon_{\rm r}$ also brings the galaxy size relation into closer agreement with the TNG L25-N512 runs, while the good agreement with observations for the black hole mass relation, the gas mass fraction and the SFRD is unaffected. Overall, the change of $\epsilon_{\rm r}$ from $0.2$ to $0.22$ leads to very minor or negligible changes in all the 6 observables plotted in Fig.~\ref{fig:L68_calibration}, especially considering the large uncertainties in the calibration data sets, and thus in principle one could do without this change.

We also considered the minimum halo mass for black hole seeding. Increasing this threshold means that, at a given time, black holes will have been growing for a shorter period of time and will consequently have a lower mass. This results in lower accretion and therefore reduces the energy released through feedback. We found that this can significantly boost star formation in higher-mass haloes (which contain larger black holes and are therefore more susceptible to black hole feedback) but has very little effect on the stellar content of low-mass haloes. We found no clear advantage to vary this in addition to the other parameters varied in this work, therefore we adjusted the black hole feedback using only the $\epsilon_{\rm r}$ parameter.

\subsubsection{Summary and further comments}

In summary, our retuned baryonic model uses updated parameter values $\rho_{\star}=0.08$, $\bar{e}_{\rm w}=0.5$ and $\epsilon_{\rm r}=0.22$, in addition to a larger gravitational softening length, to get sufficient star formation. 

While this retuning of the baryonic physics parameters has significantly boosted star formation for haloes with total mass $M_{200}\gtrsim10^{12}M_\odot$ and galaxies with stellar mass $M_\star\gtrsim5\times10^{10}M_\odot$, as shown in Fig.~\ref{fig:L68_calibration}, it is still unable to give sufficient star formation at lower masses compared to observational data. Therefore, the stellar mass fraction and stellar mass function are both underestimated at the low-mass end, and the SFRD is underestimated at redshifts $z\gtrsim5$ (when there are only low-mass haloes). We attempted to rectify this by using even lower values of $\rho_{\star}$ and reduced stellar and black hole feedback, but found that this offered little improvement overall. We even tried switching off feedback entirely, by setting the stellar wind energy to zero ($\bar{e}_{\rm w}=0$) and by preventing the seeding of black holes: while this resulted in a huge amount of star formation at masses $M_{200}\gtrsim10^{12}M_{\odot}$, there was still insufficient star formation at masses $M_{200}\lesssim10^{11.5}M_{\odot}$ to match observations. Therefore, the only way to have sufficient star formation across the full mass range appears to be by increasing the mass resolution. Interestingly, the BAHAMAS simulations \citep{McCarthy:2016mry} are able to achieve sufficient star formation for the full mass range (see the dotted lines in the top panels of Fig.~\ref{fig:L68_calibration}) despite having $\sim3\times$ lower mass resolution than our simulations. The BAHAMAS simulations were run using the \textsc{gadget-3} code \citep{Springel:2005mi}, which uses smoothed-particle hydrodynamics rather than the Voronoi mesh. Perhaps the contrasting treatments of the gas by the two codes could explain the different levels of star formation at these lowered resolutions. One possible way to further boost star formation in low-mass haloes is by having a halo mass dependency for some of the baryonic parameters, but this approach is beyond the scope of this work. On the other hand, we note that our low-resolution simulations are designed primarily for studying galaxy groups and clusters ($M_{500}\gtrsim10^{13}M_{\odot}$), for which the predictions of our model appear to be very reasonable.

While retuning these parameters, we came across a number of degeneracies. For example, as discussed above in Sec.~\ref{sec:fine_tuning:stellar_feedback}, we found that the stellar-induced wind feedback can be lowered by reducing the speed of the winds rather than the wind energy. And, in our final model, we could have instead used a slightly reduced $\rho_{\star}$ (e.g., $\rho_{\star}=0.07$) and reduced $\epsilon_{\rm r}$ (e.g., $\epsilon_{\rm r}=0.18$) to achieve similar results. Therefore, we note that different combinations of parameter values could have been used to achieve a similar level of agreement with the observational data.

\subsection{Large-box simulation}
\label{sec:methods:L302_simulations}

Our full simulation (`L302-N1136'), which has been run using the retuned baryonic model at the same mass resolution as the L68-N256 calibration runs, has a box size of $301.75h^{-1}{\rm Mpc}$ and contains $1136^3$ dark matter particles and (initially) the same number of gas cells. The simulation has been run for both GR and F5, the latter using an MG solver which has been implemented in \textsc{arepo}. This computes the highly nonlinear scalar field of $f(R)$ gravity using an adaptively refining mesh, ensuring accurate calculations of the fifth force in high-density regions \citep[for further details, see][]{Arnold:2019vpg}.

%%%%%%%%%%%%%%%%%%%%%%%%%%%%%%%%%%%%%%%%%%%%%%%%%
\begin{figure*}
\centering
\includegraphics[width=0.88\textwidth]{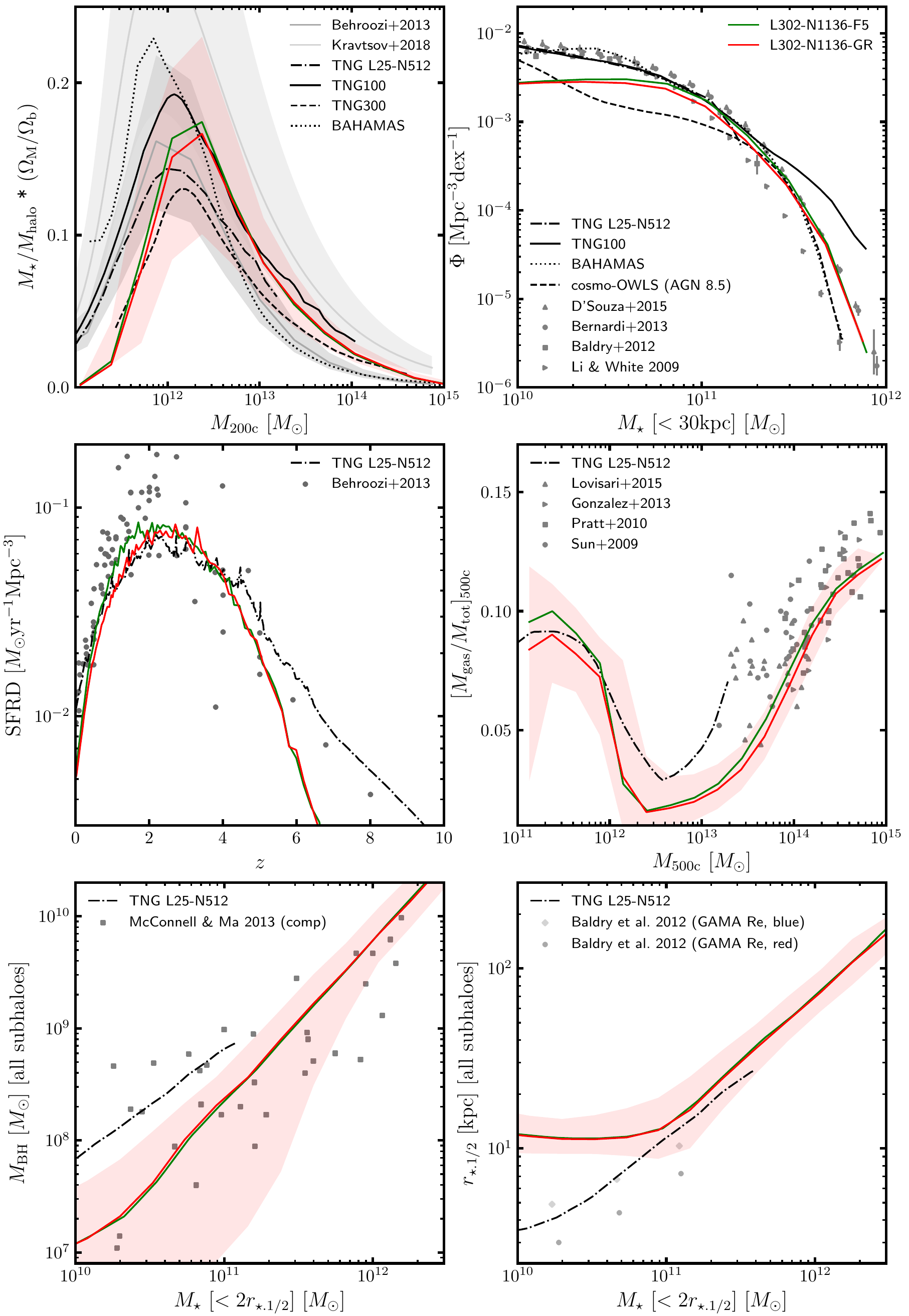}
\caption{[{\it Colour Online}] Stellar, gas and black hole properties of haloes in the L302-N1136 simulations for GR (\textit{red lines}) and F5 (\textit{green lines}). Apart from the coloured lines, the contents of this figure are identical to Fig.~\ref{fig:L68_calibration}. The red shaded regions in the subpanels for the stellar mass fraction, gas mass fraction, black-hole-mass-stellar-mass relation, and stellar-mass-galaxy-size relation indicate the 68\% spread of the GR halo data. The red shaded regions for the stellar mass function and SFRD are errors from jackknife resampling, which are barely visible.}
\label{fig:L302_observables}
\end{figure*}
%%%%%%%%%%%%%%%%%%%%%%%%%%%%%%%%%%%%%%%%%%%%%%%%%%

The red lines in Fig.~\ref{fig:L302_observables} show the GR predictions of the same six observables as used to calibrate the baryonic model. The results are slightly different compared to the cyan lines in Fig.~\ref{fig:L68_calibration}, which use the same baryonic model: the predicted amplitudes of the stellar mass fraction and stellar mass function are slightly lower, which actually improves the high-mass agreement with observations of the latter; and the amplitude of the galaxy size relation is greater for $10^{11}M_{\odot}\lesssim M_{\star}\lesssim3\times10^{11}M_{\odot}$, leading to slightly worse agreement with the TNG L25-N512 predictions at these masses. These effects are likely to be a consequence of using a much larger box size, which is less susceptible to sample variance. The L302-N1136 simulation also extends to higher masses ($M_{500}\sim10^{15}M_{\odot}$ and $M_{\star}\sim10^{12}M_{\odot}$) than the L68-N256 runs. At these masses, the agreement with the observational data in Fig.~\ref{fig:L302_observables} looks excellent. 

The predictions for the F5 model, shown by the green lines in Fig.~\ref{fig:L302_observables}, agree with the GR predictions for the galaxy size and black hole mass relations; however, the amplitudes of the other four observables are slightly boosted in F5 compared to GR. The SFRD is boosted for redshifts $0.5\lesssim z\lesssim3$; this is consistent with the results for the \textsc{shybone} simulations \citep{Arnold:2019vpg}. The stellar mass fraction and stellar mass function are boosted at $M_{200}\sim10^{12}M_{\odot}$ and $M_{\star}\sim10^{11}M_{\odot}$, respectively, and the gas mass fraction is slightly enhanced for masses $M_{500}\gtrsim10^{13}M_{\odot}$. There are a number of possible reasons for the enhanced gas fraction in F5: for example, the stronger gravitational force can lead to more gas being accumulated within $R_{500}$ by the present-day and less gas being blown away by black hole feedback. These haloes can therefore accommodate a higher level of star formation, as observed in the other panels. The F5 predictions are still in excellent agreement with the observations, therefore it is not necessary to carry out a separate retuning of the baryonic physics for this model.

\section{Results}
\label{sec:results}

Using the L302-N1136 simulations (see  Sec.~\ref{sec:methods:L302_simulations}), we have measured the observable-mass scaling relations in GR and F5 for haloes in the mass range $10^{13}M_{\odot}\leq M_{500}\lesssim10^{15}M_{\odot}$. 

In Sec.~\ref{sec:results:scaling_relations}, we will discuss the relations for the mass-weighted gas temperature $\bar{T}_{\rm gas}$, the SZ $Y$-parameter ($Y_{\rm SZ}$), the X-ray analogue of the $Y$-parameter ($Y_{\rm X}$) and the X-ray luminosity $L_{\rm X}$. These have all been computed in the same way as in \citet{Mitchell:2020aep}. The mass-weighted gas temperature is computed as follows:
\begin{equation}
    \bar{T}_{\rm gas} = \frac{\sum_i m_{{\rm gas},i}T_i}{\sum_i m_{{\rm gas},i}},
    \label{eq:mass_weighted_temperature}
\end{equation}
where $m_{{\rm gas},i}$ and $T_i$ are the mass and temperature of gas cell $i$, and the summations include all gas cells found in the radial range $0.15R_{500}<r<R_{500}$. This excludes gas cells found in the core region, which we define as the radial range $r<0.15R_{500}$, where there can be a significant dispersion in the temperature profiles due to halo mergers, black hole feedback and cooling. The $Y_{\rm SZ}$ and $Y_{\rm X}$ parameters and the X-ray luminosity have been computing using the following:
\begin{equation}
\begin{split}
    &Y_{\rm SZ} = \frac{\sigma_{\rm T}}{m_{\rm e}c^2}\sum_i N_{{\rm e},i}T_i,\\
    &Y_{\rm X} = M_{\rm gas}\times \bar{T}_{\rm gas},\\
    &L_{\rm X} = \sum_im_{{\rm gas},i}\rho_{{\rm gas},i}T_i^{1/2},
    \label{eq:observables}
\end{split}
\end{equation}
where $\sigma_{\rm T}$ is the Thomson scattering cross section, $m_{\rm e}$ is the electron rest-mass, $M_{\rm gas}$ is the total gas mass within $R_{500}$, and $N_{{\rm e},i}$ and $\rho_{{\rm gas},i}$ are the number of electrons and the density in gas cell $i$. The summations in Eq.~(\ref{eq:observables}) extend over the same radial range as for Eq.~(\ref{eq:mass_weighted_temperature}); in \citet{Mitchell:2020aep} we have checked the results of using different core region excisions, $0.1$--$0.2R_{500}$, and found $Y_{\rm SZ}$ and $Y_{\rm X}$ to be insensitive to that. 

We will test the `true density' mappings, given by Eqs.~(\ref{eq:temp_equiv_true})-(\ref{eq:lx_mapping_true}), between the F5 and GR scaling relations for redshifts $0\leq z\leq1$. We will also discuss scaling relations which do not involve the cluster mass in Sec.~\ref{sec:results:observable_relations}; these can potentially be used to test gravity using galaxy groups and clusters with no requirement to measure or infer the mass.

\subsection{Observable-mass scaling relations}
\label{sec:results:scaling_relations}

The top rows of Figs.~\ref{fig:L302_tgas}-\ref{fig:L302_lx} show the F5 and GR scaling relations for redshifts $0$, $0.5$ and $1$, with data points representing individual haloes. At $z=0$, there are $\sim8000$ GR haloes with $M_{500}>10^{13}M_{\odot}$, including $\sim500$ cluster-sized haloes with $M_{500}>10^{14}M_{\odot}$. This is a significant improvement on the L62 \textsc{shybone} simulations, which only had $\sim100$ haloes with $M_{500}>10^{13}M_{\odot}$ at $z=0$. The curves in the top rows of the figures show the median observable as a function of the mean logarithmic mass computed within mass bins; the `true density' rescalings of the F5 relations (see Eqs.~(\ref{eq:temp_equiv_true})--(\ref{eq:lx_mapping_true}), where $M_{\rm dyn}^{f(R)}/M_{\rm true}^{f(R)}$ is given by the analytical $\tanh$ formula, Eq.~\eqref{eq:mdyn_enhancement}) are indicated by the dashed lines. We use eight mass bins with constant logarithmic width over the range $10^{13}M_{\odot}\leq M_{500}\leq 10^{15.4}M_{\odot}$. All bins are shown regardless of the halo count. Although there may only be a few haloes in the highest-mass bins, we note that these correspond to high-mass clusters for which scatter in the scaling relations, especially in the model difference between F5 and GR, is expected to be very low. The relative difference between the F5 and GR binned data is shown in the lower panels of the figures.

\subsubsection{Temperature scaling relation}

%%%%%%%%%%%%%%%%%%%%%%%%%%%%%%%%%%%%%%%%%%%%%%%%%
\begin{figure*}
\centering
\includegraphics[width=0.97\textwidth]{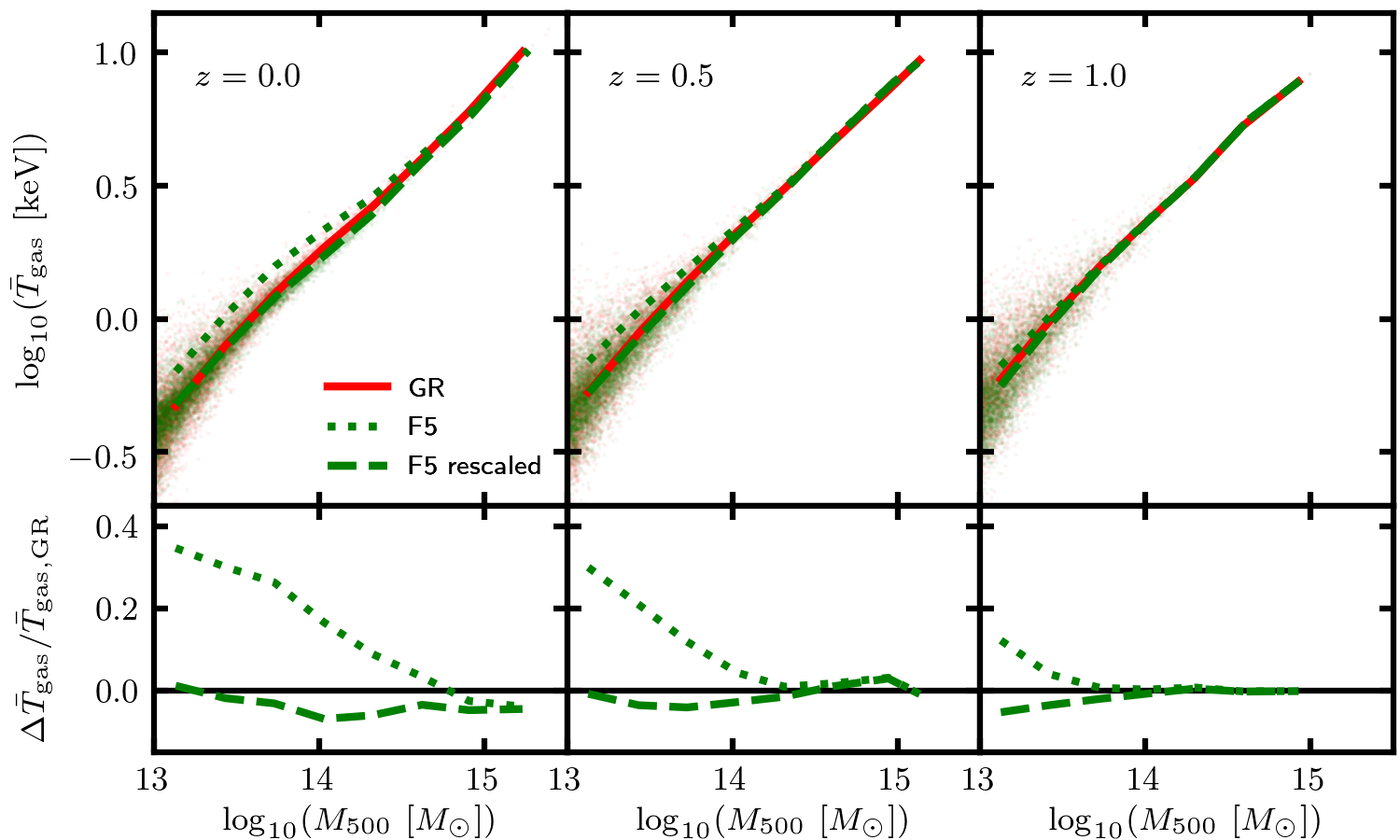}
\caption{[{\it Colour Online}] Gas temperature as a function of the halo mass for the full-physics L302 simulation (see Sec.~\ref{sec:methods}) at redshifts $0$, $0.5$ and $1$. The curves correspond to the median temperature and the mean logarithm of the halo mass $M_{500}$ computed within mass bins. Data has been included for GR (\textit{red solid lines}) and F5 (\textit{green dotted lines}). A rescaling to the F5 temperature has been carried out as described in Sec.~\ref{sec:results:scaling_relations}, as indicated by the green dashed lines. Data points are displayed, with each point corresponding to a GR halo (\textit{red points}) or to a halo in F5 (\textit{green points}), including the rescaling. \textit{Bottom row}: the relative difference between the F5 and GR curves in the above plots.}
\label{fig:L302_tgas}
\end{figure*}
%%%%%%%%%%%%%%%%%%%%%%%%%%%%%%%%%%%%%%%%%%%%%%%%%%

The results for the $\bar{T}_{\rm gas}(M)$ scaling relation are shown in Fig.~\ref{fig:L302_tgas}. The GR data appears to follow the well-known power-law behaviour for cluster-sized objects; however, the relation appears to curve at lower masses, where processes such as feedback can cause additional gas heating and break the power-law scaling. In F5, haloes are mostly screened from the fifth force for masses $M_{500}\gtrsim10^{14.5}M_{\odot}$ at $z=0$, $M_{500}\gtrsim10^{14}M_{\odot}$ at $z=0.5$ and $M_{500}\gtrsim10^{13.5}M_{\odot}$ at $z=1$, and here the F5 temperature closely follows the GR temperature. At lower masses, the F5 temperature becomes significantly enhanced, as the total gravitational potential of the halo is deepened by the fifth force. 

Our rescaling of the F5 data, which we recall involves dividing the temperature by the ratio of the dynamical mass to the true mass (Eq.~(\ref{eq:temp_equiv_true})), can successfully account for this offset at lower masses, restoring a $<7\%$ agreement with the GR relation. However, the rescaled F5 relation now slightly underestimates the GR relation on average. We note that at $z=0$, this offset appears to be roughly constant for cluster-sized masses; therefore, as long as the GR scaling relation parameters are allowed to vary in MCMC sampling (which can account for small differences in the amplitude), this rescaling is still expected to work well in our constraint pipeline presented in \citet{Mitchell:2021uzh}.

\subsubsection{\texorpdfstring{$Y_{\rm SZ}$}{YSZ} and \texorpdfstring{$Y_{\rm X}$}{YX} scaling relations}

%%%%%%%%%%%%%%%%%%%%%%%%%%%%%%%%%%%%%%%%%%%%%%%%%
\begin{figure*}
\centering
\includegraphics[width=0.97\textwidth]{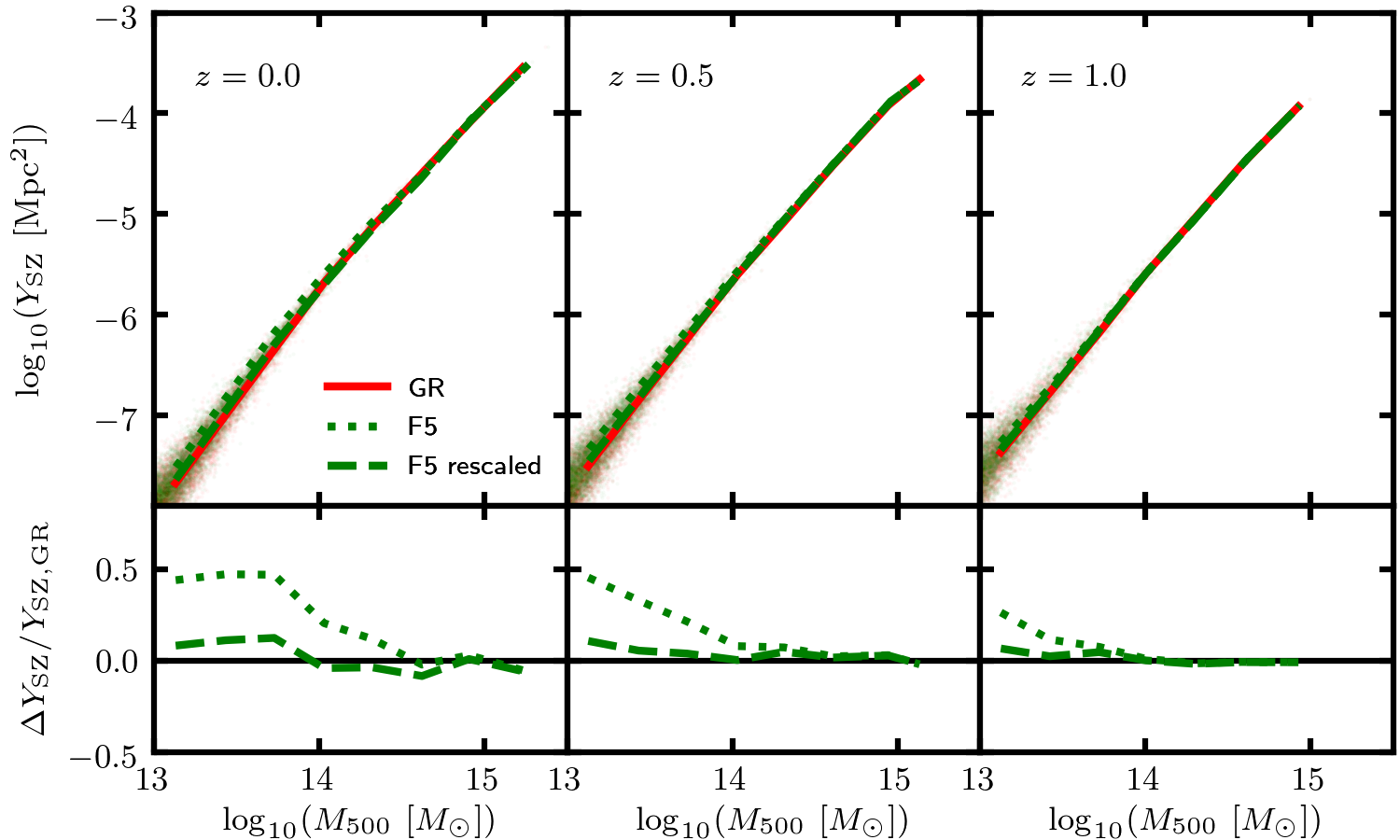}
\caption{[{\it Colour Online}] SZ Compton $Y$-parameter as a function of the halo mass for the full-physics L302 simulation (see Sec.~\ref{sec:methods}) at redshifts $0$, $0.5$ and $1$. Apart from the observable, this figure has the same layout as Fig.~\ref{fig:L302_tgas}.}
\label{fig:L302_ysz}
\end{figure*}
%%%%%%%%%%%%%%%%%%%%%%%%%%%%%%%%%%%%%%%%%%%%%%%%%%

%%%%%%%%%%%%%%%%%%%%%%%%%%%%%%%%%%%%%%%%%%%%%%%%%
\begin{figure*}
\centering
\includegraphics[width=0.97\textwidth]{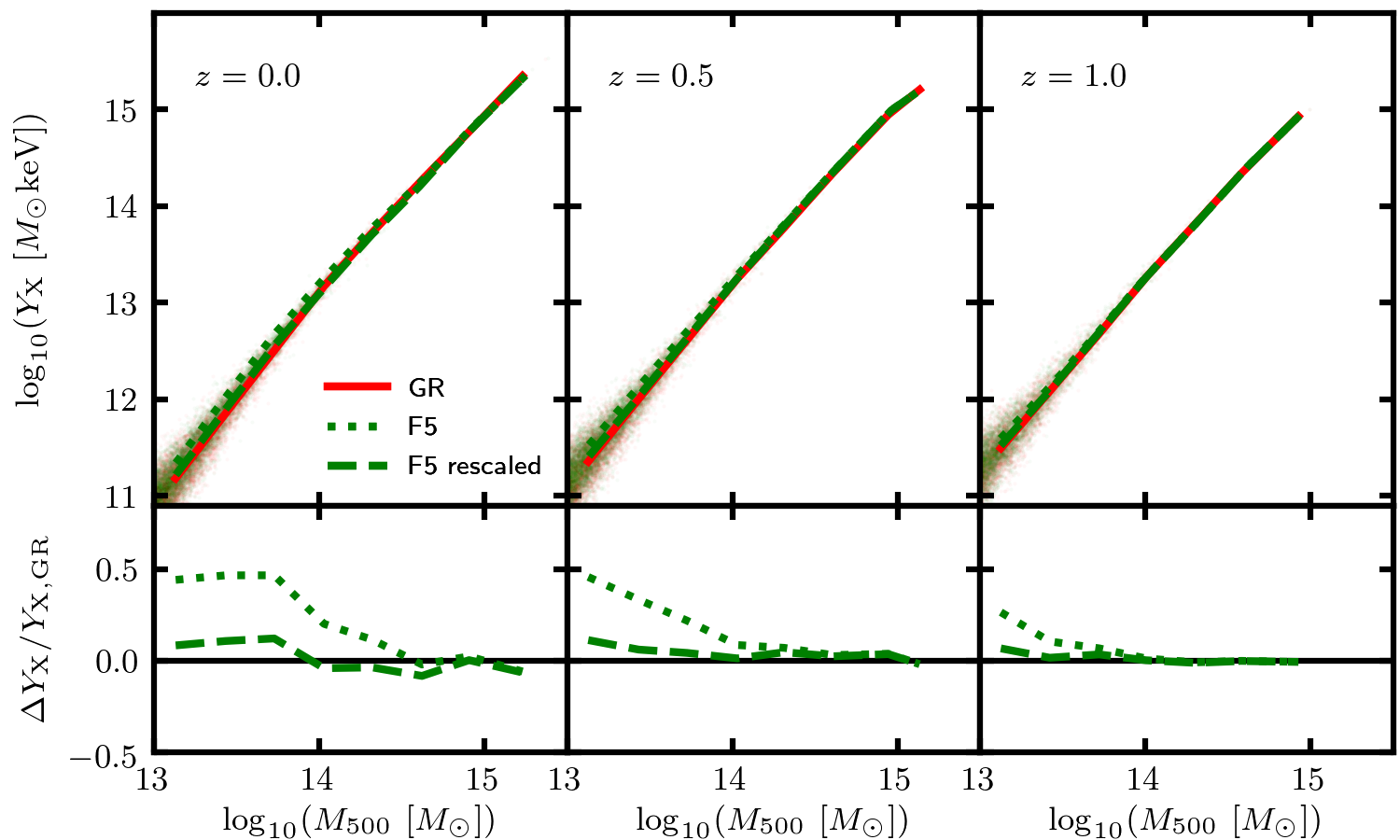}
\caption{[{\it Colour Online}] X-ray analogue $Y$-parameter as a function of the halo mass for the full-physics L302 simulation (see Sec.~\ref{sec:methods}) at redshifts $0$, $0.5$ and $1$. Apart from the observable, this figure has the same layout as Fig.~\ref{fig:L302_tgas}.}
\label{fig:L302_yx}
\end{figure*}
%%%%%%%%%%%%%%%%%%%%%%%%%%%%%%%%%%%%%%%%%%%%%%%%%%

The $Y_{\rm SZ}(M)$ and $Y_{\rm X}(M)$ relations are shown in Figs.~\ref{fig:L302_ysz} and \ref{fig:L302_yx}, respectively. The GR relation appears to follow a weakly broken power-law, with a slightly steeper slope for group-sized haloes ($M_{500}\lesssim10^{14}M_{\odot}$) than for cluster-sized haloes ($M_{500}\gtrsim10^{14}M_{\odot}$). Again the low-mass behaviour can be explained by feedback, which, in addition to heating up gas, also blows gas out from the inner regions which in turn can lower the $Y$ values. For example, in \citet{Mitchell:2020aep}, we observed that the $Y$-parameter was lower in the full-physics simulations than in the non-radiative simulations, which did not include feedback. 

For lower (unscreened) masses, we observe an enhancement of the F5 relations by up to $\sim50\%$ compared to GR. This is mostly corrected by our rescaling, after which the agreement is within $\sim12\%$ for group-sized haloes and is within a few percent on average for cluster-sized objects. This is positive news for the constraint pipeline in \citet{Mitchell:2021uzh}, which used this rescaling to model the $Y_{\rm SZ}(M)$ relation for clusters in $f(R)$ gravity in the redshift range $0<z<0.5$.

\subsubsection{X-ray luminosity scaling relation}

%%%%%%%%%%%%%%%%%%%%%%%%%%%%%%%%%%%%%%%%%%%%%%%%%
\begin{figure*}
\centering
\includegraphics[width=0.97\textwidth]{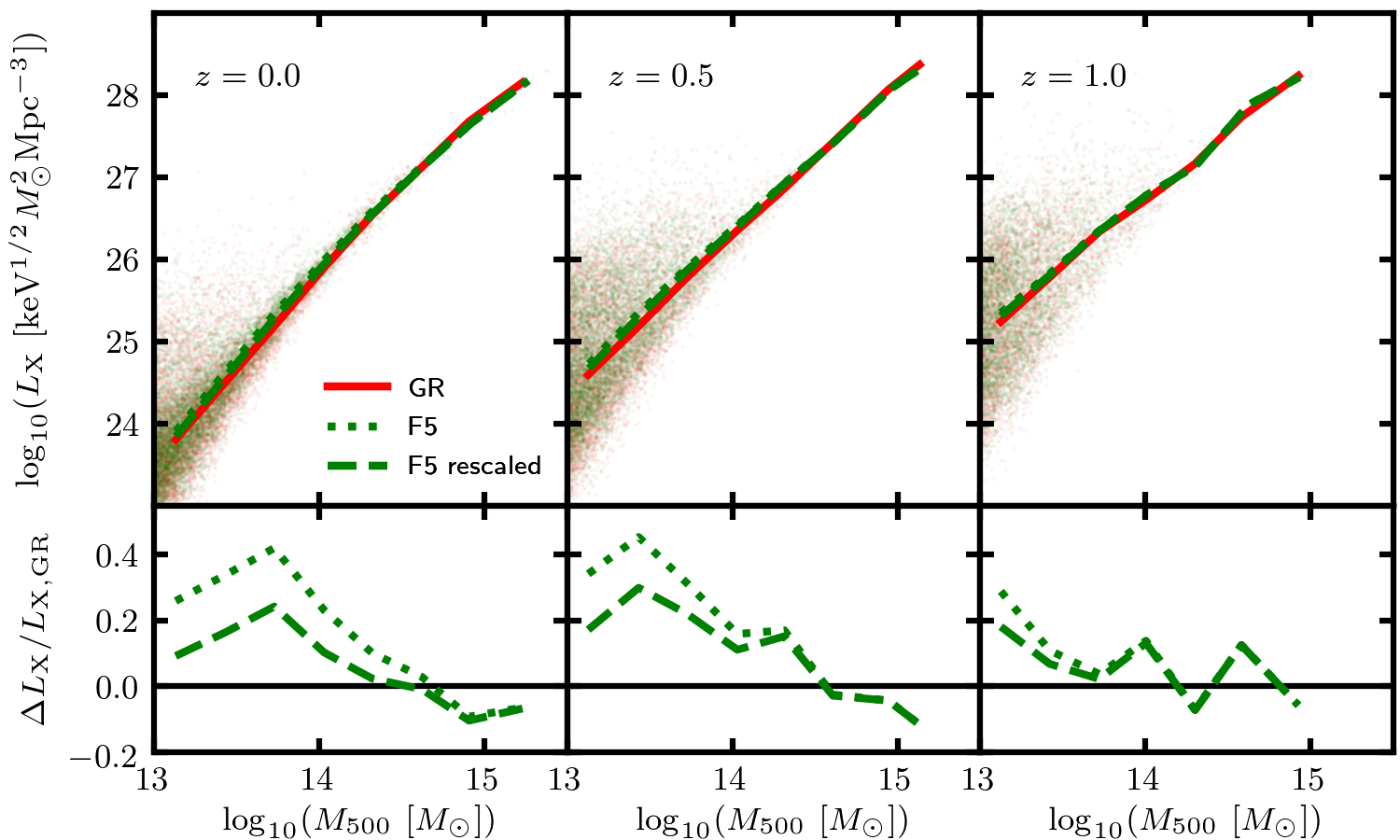}
\caption{[{\it Colour Online}] X-ray luminosity as a function of the halo mass for the full-physics L302 simulation (see Sec.~\ref{sec:methods}) at redshifts $0$, $0.5$ and $1$. Apart from the observable, this figure has the same layout as Fig.~\ref{fig:L302_tgas}.}
\label{fig:L302_lx}
\end{figure*}
%%%%%%%%%%%%%%%%%%%%%%%%%%%%%%%%%%%%%%%%%%%%%%%%%%

The $L_{\rm X}(M)$ relation is shown in Fig.~\ref{fig:L302_lx}. \citet{Mitchell:2020aep} found that our rescaling based on the ratio between the dynamical and true halo masses was unable to accurately account for the difference between GR and $f(R)$ gravity for this observable. That study was carried out primarily for group-sized haloes, and for these new results the rescaling is again unsuccessful for the mass range  $10^{13}M_{\odot}\lesssim M_{500}\lesssim10^{14}M_{\odot}$. The X-ray luminosity varies as $T_{\rm gas}^{1/2}\rho_{\rm gas}^2$. For the `true density' rescaling, which is applied here, it is assumed that the gas temperature is enhanced by the fifth force while the gas density is unchanged (see Sec.~\ref{sec:background:clusters}). This may not be a good approximation in full-physics simulations. For example, it is likely that there are different levels of feedback in F5 and GR. A greater amount of feedback in one model would result in the blowing out of gas and subsequent lowering of the gas density. This is expected to have a much greater effect on $L_{\rm X}$, which varies as $\rho_{\rm gas}^2$, than on the other observables considered in this work. For the $Y$-parameters, which vary as $T_{\rm gas}\rho_{\rm gas}$, the effects of feedback on the gas density and the temperature can roughly balance out \citep[e.g.,][]{Fabjan:2011}, allowing our rescaling to perform better for these observables as demonstrated in Figs.~\ref{fig:L302_ysz} and \ref{fig:L302_yx}. 

While the above is particularly problematic for galaxy groups, which are more susceptible to feedback, our rescaling appears to work reasonably well for cluster-sized haloes in Fig.~\ref{fig:L302_lx}, where the rescaling brings the agreement to within 10\% at $z=0$. However, the $L_{\rm X}(M)$ relation is also highly scattered compared to the other relations considered in this work. For example, the agreement between F5 and GR has a large $\sim20\%$ fluctuation at $z=1$ for $M_{500}>10^{14}M_{\odot}$, even though clusters are completely screened at this redshift.

Based on this discussion, the $\bar{T}_{\rm gas}(M)$, $Y_{\rm SZ}(M)$ and $Y_{\rm X}(M)$ relations are more suitable than the $L_{\rm X}(M)$ relation for tests of gravity that involve the cluster mass.

\subsection{\boldmath \texorpdfstring{$Y_{\rm X}$}{YX}-temperature and \texorpdfstring{$L_{\rm X}$}{LX}-temperature relations}
\label{sec:results:observable_relations}

%%%%%%%%%%%%%%%%%%%%%%%%%%%%%%%%%%%%%%%%%%%%%%%%%
\begin{figure*}
\centering
\includegraphics[width=0.97\textwidth]{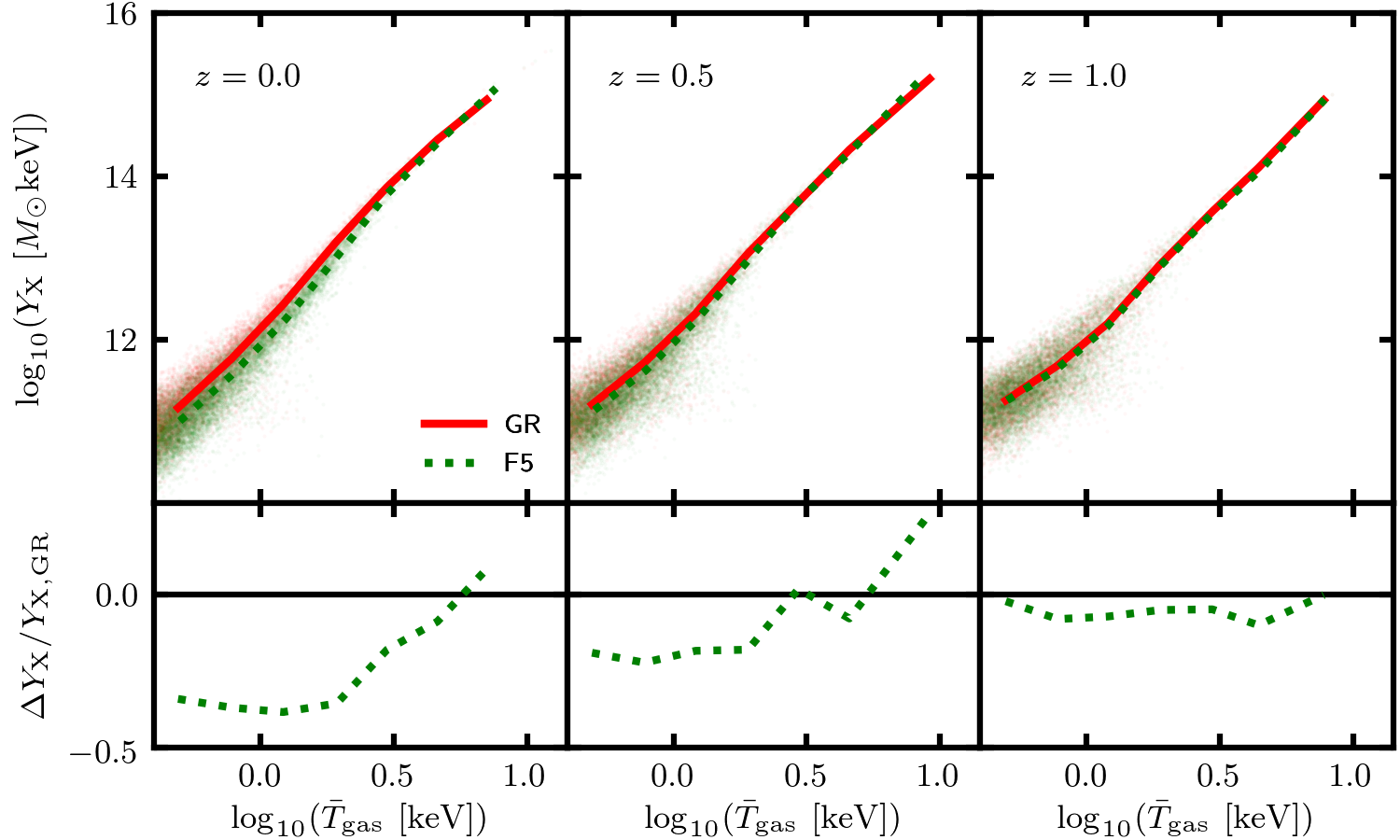}
\caption{[{\it Colour Online}] X-ray analogue of the $Y$-parameter as a function of gas temperature for haloes from the full-physics L302 simulation (see Sec.~\ref{sec:methods}) at redshifts 0, 0.5 and 1. The curves correspond to the median luminosity and the mean logarithm of the temperature computed within temperature bins. Data has been included for GR (\textit{red solid lines}) and F5 (\textit{green dotted lines}). Data points are displayed, with each point corresponding to a GR halo (\textit{red points}) or to a halo in F5 (\textit{green points}). \textit{Bottom row}: the relative difference between the F5 and GR curves in the above plots.}
\label{fig:L302_yx_tgas}
\end{figure*}
%%%%%%%%%%%%%%%%%%%%%%%%%%%%%%%%%%%%%%%%%%%%%%%%%%

%%%%%%%%%%%%%%%%%%%%%%%%%%%%%%%%%%%%%%%%%%%%%%%%%
\begin{figure*}
\centering
\includegraphics[width=0.97\textwidth]{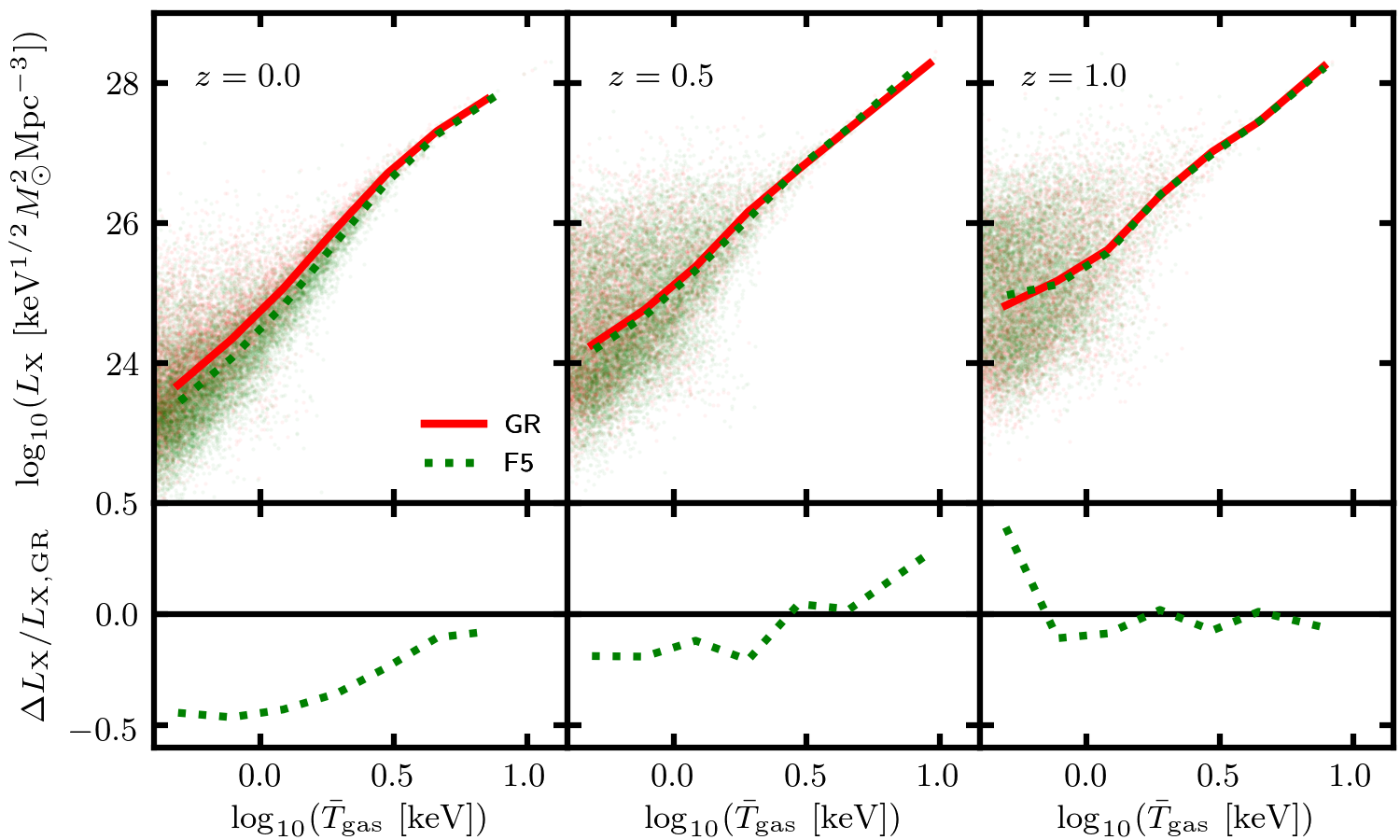}
\caption{[{\it Colour Online}] X-ray luminosity as a function of gas temperature for haloes from the full-physics L302 simulation (see Sec.~\ref{sec:methods}) at redshifts $0$, $0.5$ and $1$. Apart from the observable used in the vertical axis, this figure has the same format as Fig.~\ref{fig:L302_yx_tgas}.}
\label{fig:L302_lx_tgas}
\end{figure*}
%%%%%%%%%%%%%%%%%%%%%%%%%%%%%%%%%%%%%%%%%%%%%%%%%%

In Figs.~\ref{fig:L302_yx_tgas} and \ref{fig:L302_lx_tgas}, we show the $Y_{\rm X}(\bar{T}_{\rm gas})$ and $L_{\rm X}(\bar{T}_{\rm gas})$ relations, respectively, at redshifts 0, 0.5 and 1 (we have not shown the $Y_{\rm SZ}(\bar{T}_{\rm gas})$ relation, since this is very similar to the $Y_{\rm X}(\bar{T}_{\rm gas})$ relation). The curves show the median $Y_{\rm X}$-parameter and X-ray luminosity, respectively, and the mean logarithmic temperature computed within seven temperature bins, with logarithmic width 0.2, spanning the range $10^{-0.4}{\rm keV}\leq\bar{T}_{\rm gas}\leq10^1{\rm keV}$. 

Haloes in F5 and GR with the same temperature are expected to have a similar dynamical mass; in this case, the F5 haloes would have a lower true mass than the GR haloes, and therefore a lower gas density \citep[e.g., see the `effective density' rescalings in][]{Mitchell:2020aep}. This explains why, for $Y_{\rm X}(\bar{T}_{\rm gas})$, the amplitude of the F5 relation is suppressed by up to $\sim40\%$ compared to GR, while for $L_{\rm X}(\bar{T}_{\rm gas})$ the F5 relation is suppressed by up to $\sim45\%$ (the differences may also be partly due to differences in the levels of feedback in the two models). For both relations, the difference is greater for lower redshifts and lower temperatures, where more haloes are unscreened. 

Neither of these relations involve the cluster mass. Therefore, these could potentially be used to test gravity using galaxy groups and clusters without the risk of bias from mass measurements. This demonstrates that, besides their abundances inferred from observables such as $Y_{\rm SZ}$ and $Y_{\rm X}$, the combination of different internal observational properties for a population of galaxy clusters or groups can also offer useful, possibly complementary, constraints on the theory of gravity. We will further explore this direction in future projects.

\section{Summary, Discussion and Conclusions}
\label{sec:conclusions}

Running large-box cosmological simulations which simultaneously incorporate screened modified gravity and full baryonic physics can be computationally expensive, necessitating the use of lower mass resolutions so that the calculations can involve fewer particles. However, this means that the gas density field is smoothed, resulting in high-density peaks being lost and consequently an overall reduction in star formation. This can result in poor agreement with observations of the stellar and gaseous properties of galaxies, galaxy groups or galaxy clusters.

In this work, we have retuned the IllustrisTNG baryonic model so that it can be used to run full-physics simulations at a much lower resolution while still retaining a high level of agreement with galaxy observations. Calibrated using runs with a box size of $68h^{-1}{\rm Mpc}$, $256^3$ dark matter particles and, initially, $256^3$ gas cells, our model uses updated values for the following TNG parameters (Sec.~\ref{sec:methods:fine_tuning}): the threshold gas density for star formation, $\rho_{\star}$, is reduced from $\approx0.1$ to $0.08$; the parameter $\bar{e}_{\rm w}$ which controls the energy released by the stellar-driven wind feedback is reduced from $3.6$ to $0.5$; and the black hole radiative efficiency $\epsilon_{\rm r}$ is increased from $0.2$ to $0.22$. In addition to these changes, we have also increased the gravitational softening to a factor $1/20$ of the mean interparticle separation. By reducing the heating and blowing out of gas by feedback and two-body interactions, and lowering the threshold density of star formation, these changes boost the amount of star formation at our lowered resolution, resulting in good agreement with observations of galaxy properties including the stellar mass fraction, the stellar mass function, the SFRD and the gas mass fraction (Fig.~\ref{fig:L68_calibration}).

Using our retuned model, we have run GR and F5 simulations with a box size 301.75$h^{-1}{\rm Mpc}$ (Sec.~\ref{sec:methods:L302_simulations}). The predictions of stellar and gaseous properties in both gravity models show a very good match with galaxy observations, particularly for group- and cluster-sized masses (Fig.~\ref{fig:L302_observables}), which shows that for F5 it is not necessary to further retune the baryonic parameters. Using these simulations, we have studied, for redshifts $0\leq z\leq1$ and masses $10^{13}M_{\odot}\leq M_{500}\lesssim10^{15}M_{\odot}$, the scaling relations between the halo mass and four observable mass proxies (Sec.~\ref{sec:results:scaling_relations}): the SZ Compton $Y$-parameter $Y_{\rm SZ}$ and its X-ray analogue $Y_{\rm X}$, the mass-weighted gas temperature $\bar{T}_{\rm gas}$, and the X-ray luminosity $L_{\rm X}$.

For the $Y_{\rm SZ}(M)$ and $Y_{\rm X}(M)$ relations, our mapping between the F5 and GR relations, which involves dividing the F5 $Y$-parameter by the ratio of the dynamical mass to the true mass, is accurate to within $\sim12\%$ for galaxy groups and just a few percent for galaxy clusters. This validates our method for accounting for the effect of the fifth force on the $Y_{\rm SZ}(M)$ relation, which is currently used in our $f(R)$ constraint pipeline \citep{Mitchell:2021uzh}. For the $\bar{T}_{\rm gas}(M)$ relation, the same rescaling is again reasonable, with $\lesssim7\%$ accuracy for the full range of masses. Our rescaling does not work as well for the $L_{\rm X}(M)$ relation, which is likely due to the greater susceptibility of the X-ray luminosity to feedback processes. 

We have also shown (Sec.~\ref{sec:results:observable_relations}) that the $Y_{\rm X}$-temperature and $L_{\rm X}$-temperature scaling relations can differ in F5 and GR by up to $45\%$. These relations could potentially be used for large-scale tests of gravity that do not involve measuring the cluster mass, and hence not only eliminating one potential source of uncertainty but also including additional information in the model constraints.

By running large-box full-physics simulations for a range of $f(R)$ gravity field strengths, it will be possible to test our models for the enhancements of the dynamical mass (Eq.~(\ref{eq:mdyn_enhancement})) and the halo concentration \citep{Mitchell:2019qke} in the presence of full baryonic physics over a wide mass range. Our baryonic model can also potentially be used to run large full-physics simulations for other classes of modified gravity and dark energy models, e.g., the nDGP model \citep{Hernandez-Aguayo:2020kgq} using the MG solvers implemented in the \textsc{arepo} code, since it is likely that a recalibration of the baryonic parameters will not be necessary unless the model studied is an extreme and differs strongly from the current best-fit $\Lambda$CDM (but in that case the model is likely to have already been ruled out by other observations). The application to the nDGP model, which is another popular class of MG models, will make it possible to validate our models for the nDGP enhancements of the halo concentration and the HMF, which were calibrated using DMO simulations, in addition to extending our results for the observable-mass scaling relations to higher masses \citep{Mitchell:2021aex}. Finally, in our study of the thermal and kinetic SZ angular power spectra in $f(R)$ gravity and nDGP \citep{Mitchell:2020fnj}, we were unable to study larger angular scales ($l\lesssim500$), again due to the relatively small box size of the \textsc{shybone} simulations: this can potentially be rectified by using these larger simulations. These possibilities will be explored in future works.

The ability to run large realistic galaxy and cluster formation simulations for beyond-$\Lambda$CDM models will prove highly beneficial for research in this field: not only will this endow us with numerical tools to predict observables, such as cluster properties, that cannot be studied using DMO simulations, but the hydrodynamical simulations enabled by such a tool can be used to quantify the impacts of baryons on various other observables, such as weak lensing and galaxy clustering. The lack of such a quantitative assessment would either restrict the amount of data that can be reliably used in model tests, or lead to biased constraints on models and parameters.

\section*{Acknowledgements}

We are grateful to the IllustrisTNG collaboration for allowing us to use their baryonic physics model, and we thank Ian McCarthy, Weiguang Cui and Volker Springel for numerous helpful discussions on the model recalibration. MAM is supported by a PhD Studentship with the Durham Centre for Doctoral Training in Data Intensive Science, funded by the UK Science and Technology Facilities Council (STFC, ST/P006744/1) and Durham University. CA and BL are supported by the European Research Council via grant ERC-StG-716532-PUNCA. BL is additionally supported by STFC Consolidated Grants ST/T000244/1 and ST/P000541/1. This work used the DiRAC@Durham facility managed by the Institute for Computational Cosmology on behalf of the STFC DiRAC HPC Facility (\url{www.dirac.ac.uk}). The equipment was funded by BEIS capital funding via STFC capital grants ST/K00042X/1, ST/P002293/1, ST/R002371/1 and ST/S002502/1, Durham University and STFC operations grant ST/R000832/1. DiRAC is part of the National e-Infrastructure.

\section*{Data availability}

The simulation data used in this paper may be available upon request to the corresponding author.

%%%%%%%%%%%%%%%%%%%%%%%%%%%%%%%%%%%%%%%%%%%%%%%%%%



\bibliographystyle{mnras}
\bibliography{references} % if your bibtex file is called example.bib


%%%%%%%%%%%%%%%%%%%%%%%%%%%%%%%%%%%%%%%%%%%%%%%%%%

%%%%%%%%%%%%%%%%% APPENDICES %%%%%%%%%%%%%%%%%%%%%
\appendix

\section{Baryonic physics calibration}
\label{appendix:baryonic_fine_tuning}

In Sec.~\ref{sec:methods:fine_tuning}, we presented our new baryonic model for low-resolution, full-physics cosmological simulations. In particular, we focused on the changes that we made to the IllustrisTNG model and described only a small subset of the $\sim$200 calibration runs. In this appendix, we will provide a more detailed description of the calibration procedure, including an outline of our simulations and details of the parameter search.

Table \ref{table:simulations:fine_tuning} shows the specifications of the simulations used to tune the baryonic model. The primary goal of the tuning was to find a model that can produce sufficient star formation in low-resolution simulations to match galaxy observations. We studied in detail simulations with three different mass resolutions before settling on the resolution of the L68-N256 simulations, which have already been mentioned in Sec.~\ref{sec:methods:fine_tuning}.

%%%%%%%%%%%%%%%%%%%%%%%%%%%%%%%%%%%%%
\begin{table*}
\centering

\small
\begin{tabular}{ c@{\hskip 0.5in}cccc } 
 \toprule
 
 Specifications & \multicolumn{4}{c}{Simulations} \\
  & L100-N256 & L86-N256 & L68-N256 & L136-N512 \\

 \midrule

 box size / $h^{-1}$Mpc & 100 & 86 & 68 & 136 \\ 
 DM particle number & $256^3$ & $256^3$ & $256^3$ & $512^3$ \\
 $m_{\rm DM}$ / $h^{-1}M_{\odot}$ & $4.29\times10^9$ & $2.73\times10^9$ & $1.35\times10^9$ & $1.35\times10^9$ \\
 $m_{\rm gas}$ / $h^{-1}M_{\odot}$ & $8.3\times10^8$ & $5.3\times10^8$ & $2.6\times10^8$ & $2.6\times10^8$ \\
 number of runs & $\sim100$ & $\sim60$ & $\sim50$ & $3$ \\
 
 \bottomrule
 
\end{tabular}

\caption{Specifications of the \textsc{arepo} simulations that have been used to tune our baryonic model. These are labelled L100-N256, L86-N256, L68-N256 and L136-N512, according to their box size and dark matter particle number (we note that there are initially the same number of gas cells as dark matter particles). The simulations have all been run with standard gravity (GR).}
\label{table:simulations:fine_tuning}

\end{table*}
%%%%%%%%%%%%%%%%%%%%%%%%%%%%%%%%%%%%%

\subsection{L100-N256 simulations}

To start with, we used simulations with a box size of $100h^{-1}{\rm Mpc}$, containing $256^3$ dark matter particles and (initially) the same number of gas cells (L100-N256). With an average gas cell mass of $\sim8.3\times10^8h^{-1}M_{\odot}$, these have 512 times lower mass resolution than the simulations used to calibrate the fiducial TNG model and the same resolution as the BAHAMAS simulations \citep{McCarthy:2016mry}, which were run using \textsc{gadget-3} \citep{Springel:2005mi} rather than \textsc{arepo}. We ran $\sim$100 simulations at this resolution, varying the following baryonic parameters: the threshold gas density for star formation $\rho_{\star}$ (see Sec.~\ref{sec:methods:fine_tuning:star_formation_model}) was varied in the range $[0.00,0.13]~{\rm cm^{-3}}$; the parameter $\bar{e}_{\rm w}$ controlling the stellar wind energy (see Eq.~(\ref{eq:wind_energy})) was varied in the range $[0.0,3.6]$; the parameters $\kappa_{\rm w}$ and $v_{\rm w,min}$ controlling the stellar wind speed (see Eq.~(\ref{eq:wind_speed})) were varied over ranges $[0.0,29.6]$ and $[0,500]~{\rm kms^{-1}}$, respectively; and the black hole radiative efficiency $\epsilon_{\rm r}$ (see Sec.~\ref{sec:methods:fine_tuning:bh_feedback}) was varied in the range [0.02,0.20]. We also tested gravitational softening lengths in the range 1/40 to 1/10 times the mean inter-particle separation.

These runs provided a very useful insight into the effects of changing each parameter, however all of the tested parameter combinations at the L100-N256 resolution resulted in insufficient star formation within haloes of mass $M_{200}\lesssim10^{13}M_{\odot}$, and at higher halo masses it was difficult to simultaneously match observations for different galaxy properties. For example, parameter combinations which yielded a sufficiently high stellar mass function typically resulted in the stellar mass fraction being overestimated, and in order to match the SFRD observations it was necessary to set either $\rho_{\star}$ or the stellar wind energy close to zero. We therefore decided to look at higher resolutions.

\subsection{L86-N256 simulations}

Keeping the dark matter particle number (and initial gas cell number) unchanged, we initially reduced the box size to $86h^{-1}{\rm Mpc}$ (L86-N256), and executed $\sim$60 runs at this higher resolution. Our best model used a gravitational softening length of 1/20 times the mean inter-particle separation and the following parameter combination: $\rho_{\star}=0.05{\rm cm^{-3}}$, $\bar{e}_{\rm w}=0.5$, $\kappa_{\rm w}=2$, $v_{\rm w,min}=200~{\rm kms^{-1}}$ and $\epsilon_{\rm r}=0.15$. This gave a reasonable match with high-mass observations of the stellar mass fraction ($M_{200}\gtrsim2\times10^{12}h^{-1}M_{\odot}$) and stellar mass function ($M_{\star}\gtrsim10^{11}h^{-1}M_{\odot}$), however the agreement was still poor at lower masses and the SFRD was significantly underestimated for redshifts $z\gtrsim2$. We made some efforts to rectify this. For example, we switched off feedback entirely by setting the wind energy to zero and preventing the formation of black hole particles, and we tried using much lower values of $\rho_{\star}$. While these efforts resulted in more star formation at lower masses, it was still not enough to match observational data, and at higher masses and lower redshifts there was now far too much star formation.

\subsection{L68-N256 and L136-N512 simulations}

We finally settled on the $68h^{-1}{\rm Mpc}$ (L68-N256) box size, where we ran a further $\sim$50 runs to calibrate the final model presented in Sec.~\ref{sec:methods:fine_tuning}. Our final model, with $\rho_{\star}=0.08{\rm cm^{-3}}$, $\bar{e}_{\rm w}=0.5$, $\epsilon_{\rm r}=0.22$ and a softening length of 1/20 times the mean inter-particle separation, is able to produce sufficient star formation at lower masses and higher redshifts than the above L86-N256 model, and only requires changes to three of the TNG model parameters (we take the TNG values for $\kappa_{\rm w}$ and $v_{\rm w,min}$). The predictions from five of the L68-N256 runs are shown in Fig.~\ref{fig:L68_calibration} to illustrate the effects of each of the changes to the fiducial TNG parameter values.

In order to assess the effects of sample variance, we also ran three simulations with an increased box size of $136h^{-1}{\rm Mpc}$ (L136-N512) and the same mass resolution as the L68-N256 runs. The results from these simulations, which were run using our three most promising baryonic models, indicated that the stellar mass fraction and stellar mass function are slightly reduced in the larger box. This is why we selected the above model, even though it slightly overestimates the stellar mass function in Fig.~\ref{fig:L68_calibration}.

%%%%%%%%%%%%%%%%%%%%%%%%%%%%%%%%%%%%%%%%%%%%%%%%%%

% Don't change these lines
\bsp	% typesetting comment
\label{lastpage}
\end{document}